\documentclass[aps,prx,showpacs]{revtex4}

\usepackage{graphicx}
\usepackage{hyperref}
\usepackage{amsfonts}
\usepackage{amsbsy}
\usepackage{amssymb}
\usepackage{xcolor}
\usepackage[normalem]{ulem}
 \usepackage{float}
 \usepackage{multirow}

\begin{document}

\preprint{}

\title{Dynamical robustness of discrete conservative systems: \\
Harper and generalized standard maps}

\author{Ugur Tirnakli$^{1}$}
\email{ugur.tirnakli@ege.edu.tr}
\author{Constantino Tsallis$^{2,3,4}$}
\email{tsallis@cbpf.br}
\author{Kivanc Cetin$^{1,5}$}
\email{kivanccetin@hotmail.com}
\affiliation{ $^1$Department of Physics, Faculty of Science, Ege University, 35100 Izmir, Turkey \\
$^2$Centro Brasileiro de Pesquisas Fisicas and National Institute of Science and Technology for Complex Systems \\ 
\mbox{Rua Xavier Sigaud 150, Rio de Janeiro 22290-180, Brazil}\\
$^3$ Santa Fe Institute, 1399 Hyde Park Road, Santa Fe, New Mexico 87501, USA \\
$^4$ Complexity Science Hub Vienna, Josefst\"adter Strasse 39, 1080 Vienna, Austria \\
$^5$ American Collegiate Institute, 35290 Izmir, Turkey
 }

\date{\today}

\begin{abstract}
In recent years, statistical characterization of the discrete conservative dynamical systems 
(more precisely, paradigmatic examples of area-preserving maps such as the standard and the web 
maps) has been analyzed extensively and shown that, for larger parameter values for which  the 
Lyapunov exponents are largely positive over the entire phase space, the probability distribution is a 
Gaussian, consistent with Boltzmann-Gibbs (BG) statistics. 
On the other hand, for smaller parameter values for which the Lyapunov exponents are virtually zero 
over the entire phase space, we verify this distribution appears to approach a $q$-Gaussian 
(with $q= 1.935\pm0.005$), consistent with $q$-statistics. 
Interestingly, if the parameter values are in between these two extremes, then the probability 
distributions happen to exhibit a linear combination of these two behaviours.
Here, we numerically show that the Harper map is also in the same universality class of the maps 
discussed so far. This constitutes one more evidence on the robustness of this behavior whenever 
the phase space consists of stable orbits.
Then, we propose a generalization of the standard map for which the phase space includes many 
sticky regions, changing the previously observed simple linear combination behavior to a more 
complex combination. 
\end{abstract}

\maketitle

\section{Introduction}
\label{sec:1}

The dynamics of the ergodic or mixing systems can be explained via notions of the Boltzmann-Gibbs (BG) 
statistical mechanics. For these systems, exponential and Gaussian distributions are appropriate forms 
to describe the limit probability behavior of relevant variables of the system under consideration. 
Since these distributions maximize the BG entropy, the reason behind the occurrence of such distributions 
can be explained by the Central Limit Theorem (CLT). On the other hand, due to the ergodicity breakdown 
and strong correlations among the random variables observed for some systems for some intervals of 
system parameters, BG statistical mechanical approaches fail to describe the dynamics of those cases 
and the CLT is not valid anymore. It has been shown in recent years that a generalization of the CLT 
is possible and the limit probability distributions seem to converge to a $q$-Gaussian 
distribution \cite{Umarov1,Umarov2,Nelson,Vignat,Umarov3,Hahn} for a class of systems with certain 
correlations. Like the role of the Gaussians in the BG statistics, $q$-Gaussian distributions maximize 
the nonadditive entropy ($S_q=k(1-\sum_i p_i^q)/(q-1)$) and constitute the basis of nonextensive statistical 
mechanics \cite{Tsallis,Tsallis2010}. The nonextensive statistical mechanical framework provides a 
more general picture by recovering the BG statistical framework as a special case ($q \rightarrow 1$) 
and this generalization seems to be the appropriate tool for explaining the statistical mechanical 
behavior of a wide range of systems where BG statistics is known to fail.

In recent years, domains of validity of these two statistical mechanical frameworks have been shown 
by utilizing the rich phase space behaviors exhibited in area-preserving maps \cite{Ugur,Ruiz,Ruiz2017b} 
(namely, the standard map and the web map). As the amount of nonintegrability increases with the 
increment of the map parameter for these Hamiltonian systems, chaotic and regular behavior regions may 
coexist in the phase spaces of these maps for specific parameter values. In the chaotic regions, 
the system is ergodic and iterates of the chaotic trajectories display uncorrelated behavior while 
wandering throughout the allowed region in the phase space. In this case, limit probability distribution 
of sums of iterates of the map converges to a Gaussian consistently with assertions of the BG statistical 
mechanical framework. On the other hand, iterates of trajectories starting from initial conditions 
located inside the nonergodic stability islands exhibit strongly correlated behavior. For these initial 
conditions, it has been shown for the standard map \cite{Ugur,Ruiz} and the web map \cite{Ruiz2017b} 
that the limit probability distribution converges to a $q$-Gaussian with a specific $q= 1.935\pm0.005$ 
value. When the entire system is modeled by using initial conditions coming from both chaotic sea and 
stability islands for some specific parameter values, the limit probability distribution is obtained as 
a linear combination of a Gaussian and a $q$-Gaussian with $q= 1.935\pm0.005$; the $q$-Gaussian distribution 
maintains its existence together with Gaussian even for large number of iteration steps. This limit 
distribution seems very robust from the statistical mechanical point of view since the same 
distribution appears for the initial conditions selected from stability islands of different maps. 
Any novel understanding of such systems would no doubt be important if we consider the role of the 
area-preserving maps in physics and in the development of the chaos theory \cite{Hilborn}. 
Many physical systems such as magnetic traps \cite{Chirikov1987}, electron magnetotransport 
in classical and quantum wells \cite{Shepelyansky}, particle accelerators \cite{Izraelev} can be modeled 
by using the standard map as a first approximation. In addition, linear combination of the standard map 
and the web map can be used for modeling many physical systems \cite{Zaslavsky91}. 
As the area-preserving maps have such importance in different branches, whether the common limit 
probability behavior observed for initial conditions of the stability islands of the standard map and 
the web map is a universal behavior for all area-preserving maps would be a very intriguing research 
question that deserves to be investigated. In this study, to test the robustness of the $q$-Gaussian 
distribution with $q= 1.935\pm0.005$, we analyze the limit behavior of the sums of the iterates of the 
Harper map \cite{harper}, which models transport phenomena in deterministic chaotic Hamiltonian systems. 
In addition to the Harper map, we also define a new generalized form for the standard map, 
namely $z$-generalized standard map, to create independent and unique area-preserving 
maps exhibiting different phase space dynamics.

In order to statistically characterize these area-preserving maps we choose various map parameter 
values for each system where the phase spaces display different behavior. As the phase spaces of 
these scenarios provide rich observations by exhibiting chaotic and regular behaviors at the same time 
for specific parameter values, we numerically investigate the limit probability distribution of the 
entire system and visualize how the limit distributions vary according to the map parameter value. 
For each case we obtain phase space portraits given in this paper by iterating $40$-$50$ randomly 
chosen initial conditions $T=5\times 10^{3}$ times. In order to quantify the trajectory behaviors seen 
in the phase portraits we calculate the largest Lyapunov exponent, LLE ($\lambda$), by using Benettin 
algorithm \cite{Benettin86} for each initial condition randomly chosen from the entire phase space. 
Lyapunov exponents are calculated over $T=5\times 10^{5}$ time steps using $M=5\times 10^{5}$ 
initial conditions and Lyapunov spectra of scenarios are portrayed via color maps in order to reveal the 
regions with different behavior. Since the chaotic regions and the stability islands exhibit largely 
positive and nearly zero LLE values respectively, this separation of the phase space regions allows 
us to distinguish the portions of the phase space where the system appears to be ergodic and 
nonergodic \cite{Ugur}. Chaotic trajectories presenting largely positive LLE values diverge 
exponentially in the allowed region of the phase space and these trajectories spread into this region 
with apparently random behavior. For the strongly chaotic regions, system exhibits mixing property 
and ergodic behavior. On the other hand, trajectories located inside the stability islands, which can 
be periodic or quasi-periodic, present nearly zero LLE values ($\lambda\approx0$) and the system 
is nonergodic in these regions. Although argument about the ergodicity of the chaotic trajectories is 
verified for numerous dissipative \cite{Afsar2013,Cetin2015} and conservative \cite{Ugur,Ruiz,Ruiz2017b} 
systems, we come across with a contrary situation for sticky chaotic regions that may arise in 
several $z$-generalized standard map systems by exhibiting nonergodic behavior for finite observation 
times. This unexpected behavior will be further discussed later on, when analyzing the $z$-generalized 
standard map.

Since the ergodic and nonergodic portions of the phase space require different statistical mechanical 
approaches, we can investigate the limit probability distributions of the variables of the systems to 
determine the domains of validity of the BG and of the nonextensive statistical mechanics. 
In the spirit of the Central Limit Theorem, for the limit probability distribution characterization, 
we define the variable
\begin{equation}
	y= \frac{1}{\sqrt{T}} \sum_{i=1}^{T} \left(x_i- \langle \, x \, \rangle \right) \label{eq:variable}
\end{equation}
where $x$ is the variable of the map, $\langle \, \cdots \rangle$ denotes averaging over a large 
number of iterations $T$ and a large number of randomly chosen initial conditions $M$, i.e., 
$\langle x \rangle=\frac{1}{M}\frac{1}{T}\sum_{j=1}^{M}\sum_{i=1}^T x_i^{(j)}$. 
It was previously shown, for arbitrary values of the parameter $K$ of the standard map \cite{Ugur,Ruiz}, 
that the probability distribution of these sums (Eq.~(\ref{eq:variable})) can be modeled as
\begin{equation}
	\label{eq:qGauss}
	P_q(y;\mu_q,\sigma_q)=A_q\sqrt{B_q}\left[1-(1-q)B_q(y-\mu_q)^2 \right]^{\frac{1}{1-q}},
\end{equation}
that represents the probability density for the initial conditions inside the vanishing Lyapunov region 
($q\ne1$), where $\mu_q$ is the $q$-mean value, $\sigma_q$ is the $q$-variance, $A_q$ is the 
normalization factor and $B_q$ is a parameter which characterizes the width of the 
distribution \cite{prato-tsallis-1999}:

\begin{equation}
	\label{eq:Aq}
	A_q=\left\{\begin{array}{lc}\displaystyle\frac{\Gamma\left[\frac{5-3q}{2(1-q)}\right]}
		{\Gamma\left[\frac{2-q}{1-q}\right]}\sqrt{\frac{1-q}{\pi}}, &q<1\\
		\displaystyle\frac{1}{\sqrt{ \pi }},&q=1\\
		\displaystyle\frac{\Gamma \left[\frac{1}{q-1}\right]}{\Gamma\left[\frac{3-q}{2(q-1)}\right]} 
		\sqrt{\frac{q-1}{\pi}}, &1<q<3
	\end{array} \right.
\end{equation}

\begin{equation}
	\label{eq:Bq}
	B_q=[(3-q)\sigma_q^2]^{-1} .
\end{equation}
The $q$-mean value and $q$-variance are defined by (see \cite{prato-tsallis-1999} for the 
continuous version):
\begin{equation}
	\mu_q = \frac{ \sum_{i=1}^N y_i [P_q(y_i)]^q }
	{ \sum_{i=1}^N [P_q(y_i)]^q },
\end{equation}

\begin{equation}
	\sigma_q^2 = \frac{ \sum_{i=1}^N y_i^2 [P_q(y_i)]^q }
	{ \sum_{i=1}^N [P_q(y_i)]^q },
\end{equation}
though we have considered these variables as fitting parameters.

The $q \to 1$ limit recovers the Gaussian distribution 
$P_1(y; \mu_1, \sigma_1)=\frac{1}{\sigma_1 \sqrt{2 \pi }} 
\exp\left[-\frac{1}{2}\left(\frac{y-\mu_1}{\sigma_1} \right)^2\right]$. 
In the analyses of the limit probability distributions of such maps with various values for the map 
parameter, we randomly choose a large number of initial conditions, larger than $3\times 10^{7}$,   
from the entire phase space, and numerically calculate the limit distribution of Eq.~\ref{eq:variable} 
using $T=2^{22}$ iteration steps in order to obtain a good statistical description of the systems. 
These values are determined in accordance with the recent works \cite{Ugur,Ruiz} where they have 
been shown to be optimal by considering required computational times and convergence of the obtained 
probability distributions.

This paper is organized as follows: 
Firstly, the Harper map and the $z$-generalized standard map will be introduced in Sections II and III; 
then the results of the numerical calculations are discussed in Section IV. 
Finally we conclude in the last section.

\section{Harper Map }

In order to investigate transport phenomena in deterministic chaotic Hamiltonian 
dynamics, a two-dimensional area-preserving map has been proposed in \cite{harper} 
using a time-dependent Hamiltonian of the form 

\begin{equation}
\begin{array}{l}
H(x,p,t) = -V_2 \cos(2\pi p) - V_1 \cos(2\pi x) \tau \sum_{n=-\infty}^{\infty} 
\delta(t-n\tau) .
\end{array}
\label{eq:harper}
\end{equation}
\noindent This is called kicked Harper model and has several applications in 
physics \cite{harperApp1,harperApp2,harperApp3}.
If one integrate Eq.(\ref{eq:harper}) over one period $\tau$ of the kicking 
potential, the kicked Harper map can be obtained easily as follows: 

\begin{equation}
\begin{array}{l}
p_{i+1} = p_i - \gamma_1 \sin(2\pi x_i)\\
 x_{i+1}=x_i + \gamma_2 \sin(2\pi p_{i+1})
\end{array}
\label{eq:Harpermap}
\end{equation}
where $p$ and $x$ are taken as modulo $1$ and $\gamma_j=2\pi V_j \tau$. 
In the present paper, we will only consider the special case $\gamma_1=\gamma_2 \equiv \gamma$.

\section{Generalized Standard Map}

The Hamiltonian of the standard map is given by \cite{Zaslavsky2007}
\begin{equation}
H = \frac{1}{2} p^2 - K \cos(x) \sum_{n=-\infty}^{\infty}\delta\left(\frac{t}{T}-n\right)\\
\label{eq:hamiltonian}
\end{equation}
where $p$ and $x$ are the momentum and position of a particle respectively and the periodic sequence of 
$\delta$-pulses has the period $T=2\pi/\nu$. 
The equations of motion derived from this Hamiltonian enables us to write down the momentum and position 
variables at the $n$-th and ($n+1$)-th kicks, from where the original standard map is 
obtained \cite{Chirikov1979}. 
One possible way of generalizing the original standard map is to modify the kicked term as 
$(K/z) \cos(zx)$, which results in defining the $z$-generalized standard map as follows:
\begin{equation}
\begin{array}{l}
p_{i+1} = p_i - K \sin(z x_i)\\
 x_{i+1}=x_i+p_{i+1}
\end{array}
\label{eq:smap}
\end{equation}
where $p$ and $x$ are taken as modulo $2\pi$, $K$ is the map parameter which controls the 
amount of nonintegrability of the system and $z$ is an integer; $z=1$ recovers the usual standard map. 
With different $z$ values we define unique systems exhibiting specific phase-space dynamics. 
In this paper we investigate the phase-space behaviors and the limit distributions for various $z$ 
values with $K=0.2$ and $K=0.6$ parameters by using the numerical calculations introduced in the 
Introduction section. In order to present a clear evolution for the phase portraits and the limit 
distributions according to the $z$ generalization term, we analyze $z=3$, $z=5$, $z=40$ generalized 
systems for $K=0.2$, and $z=3$, $z=4$, $z=15$ systems for $K=0.6$. For both parameter values we also 
analyze the $z=1$ system (the original standard map) in order to study how the $z$-generalization 
modifies the dynamical and statistical behavior of the system.

\section{Results}

\begin{figure}[H]
\centering
\includegraphics[height=5cm,angle=0]{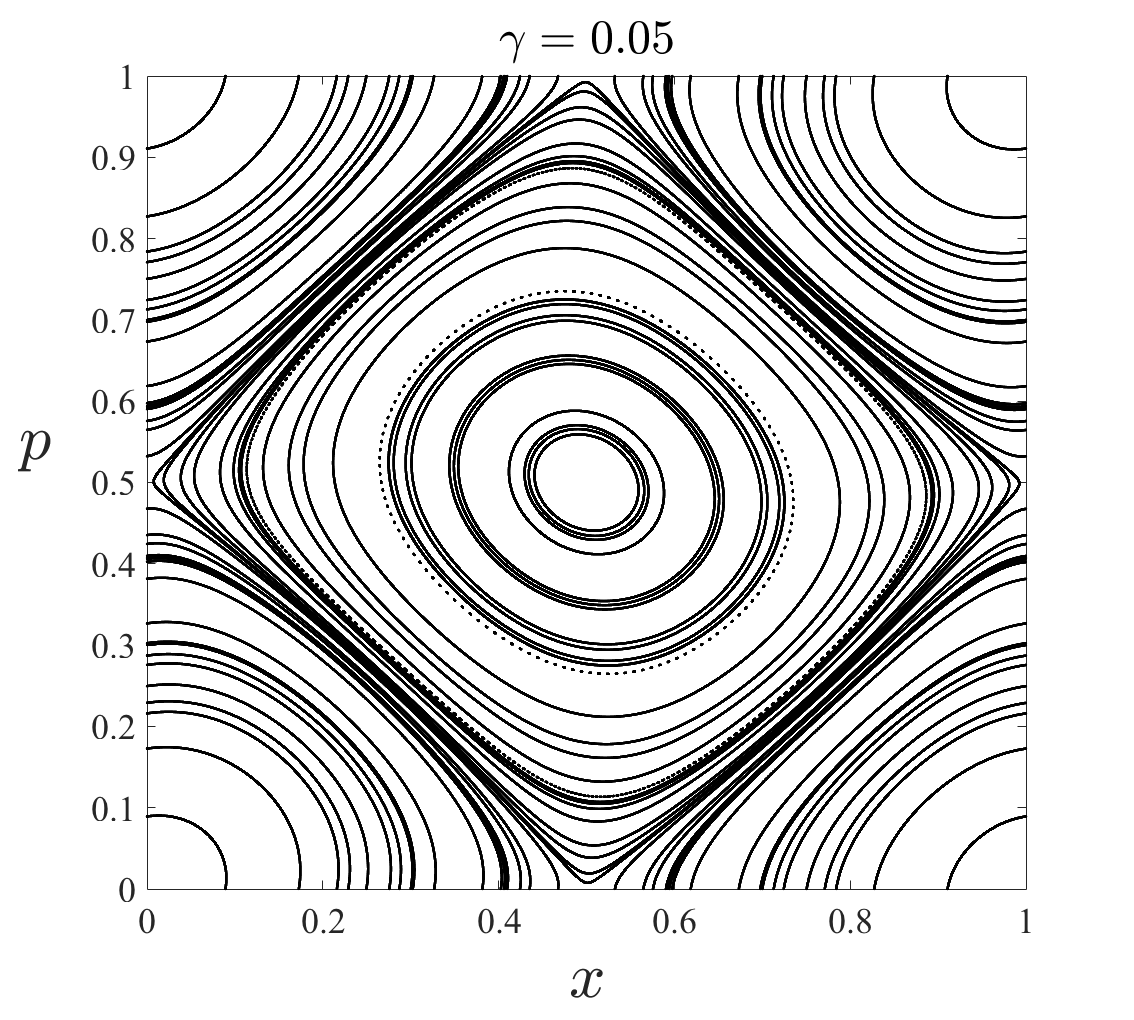}
\includegraphics[height=5cm,angle=0]{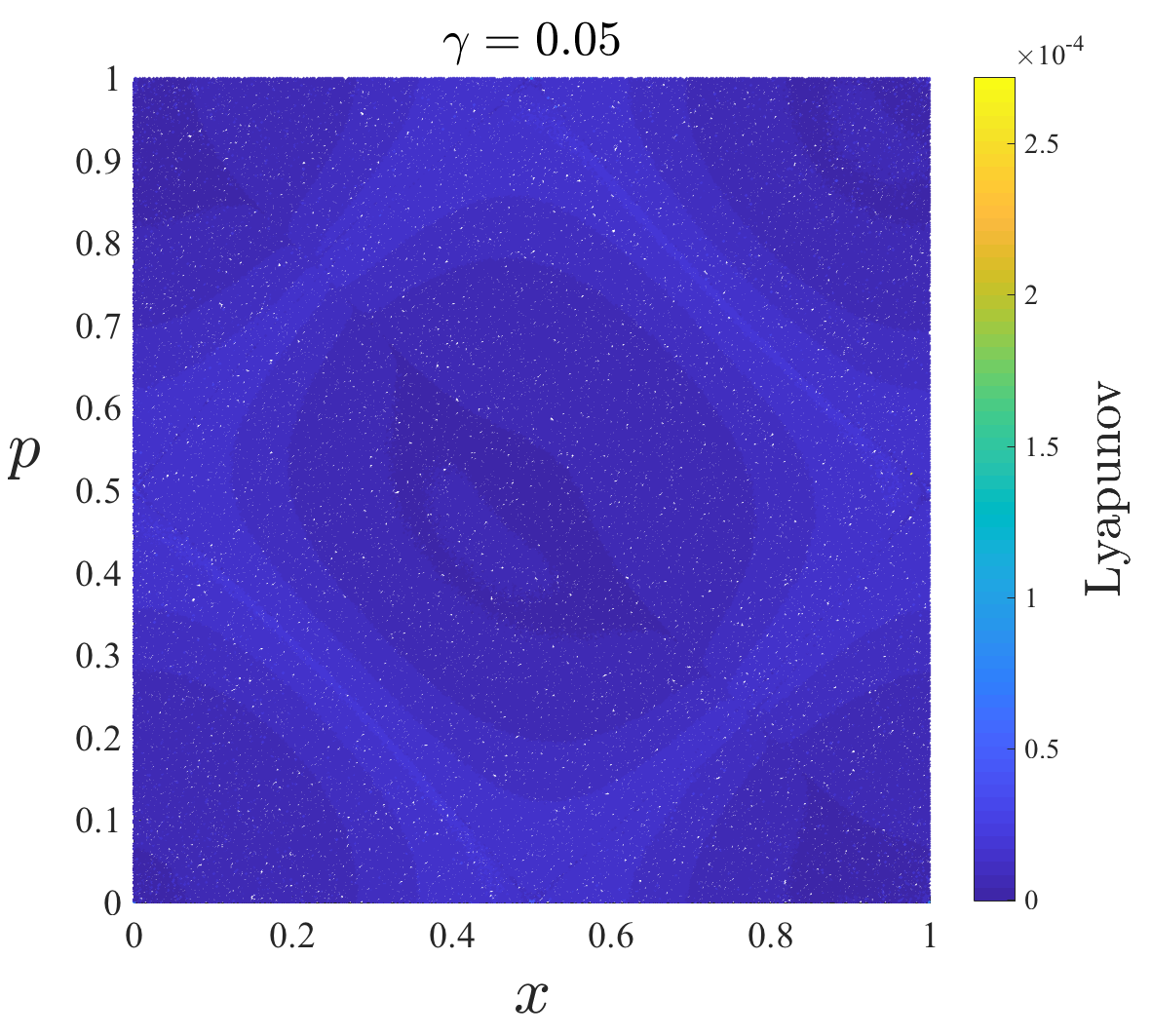}
\includegraphics[height=4.8cm,angle=0]{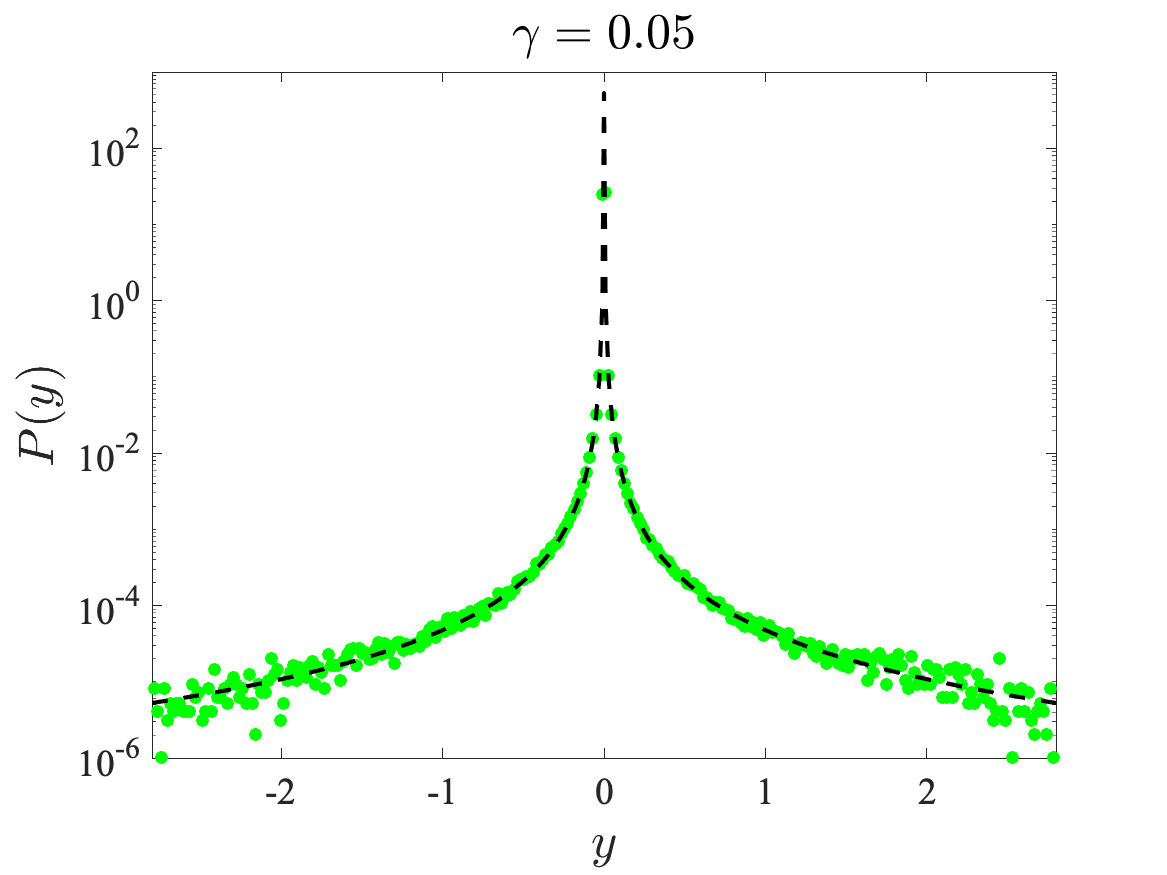}\\
\includegraphics[height=5cm,angle=0]{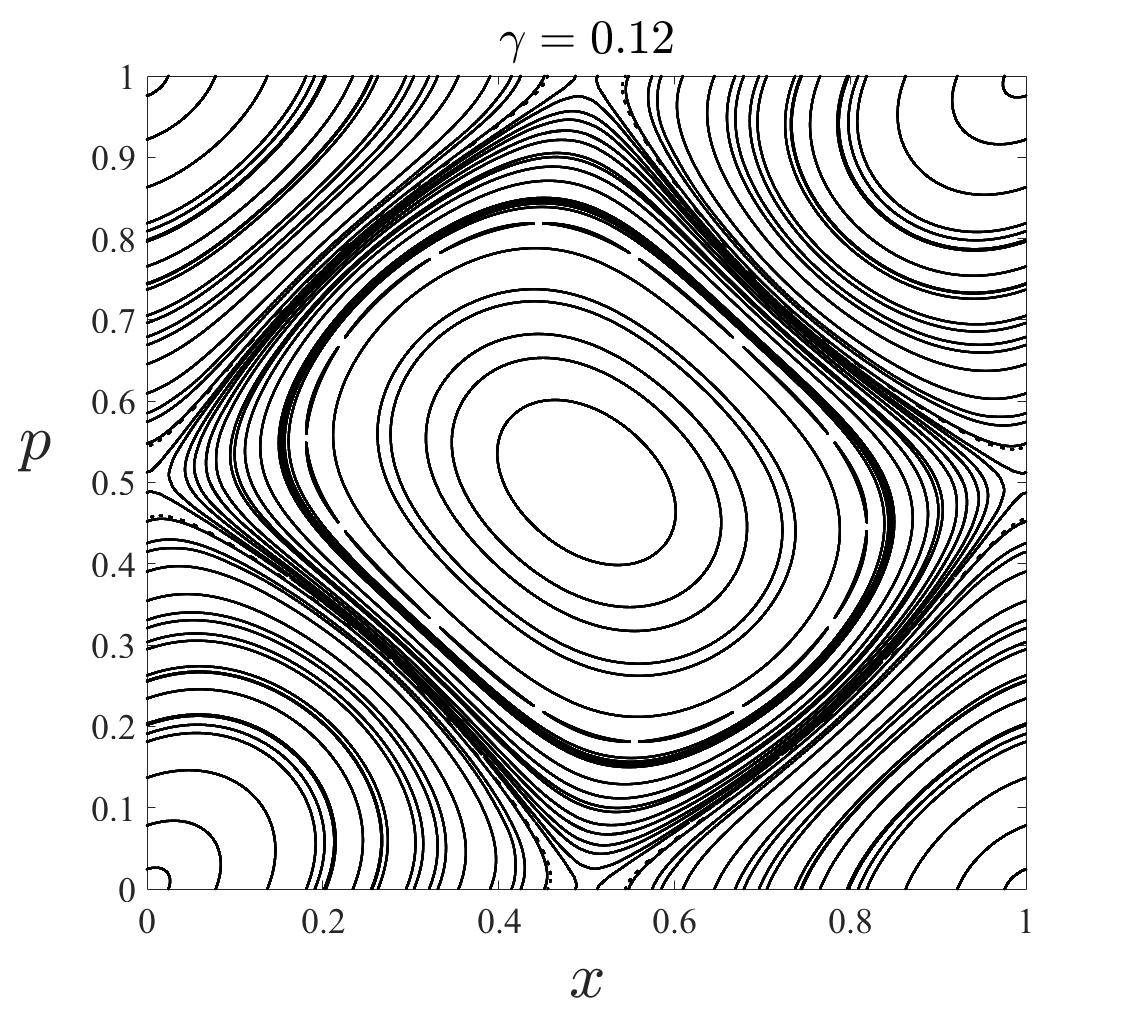}
\includegraphics[height=5cm,angle=0]{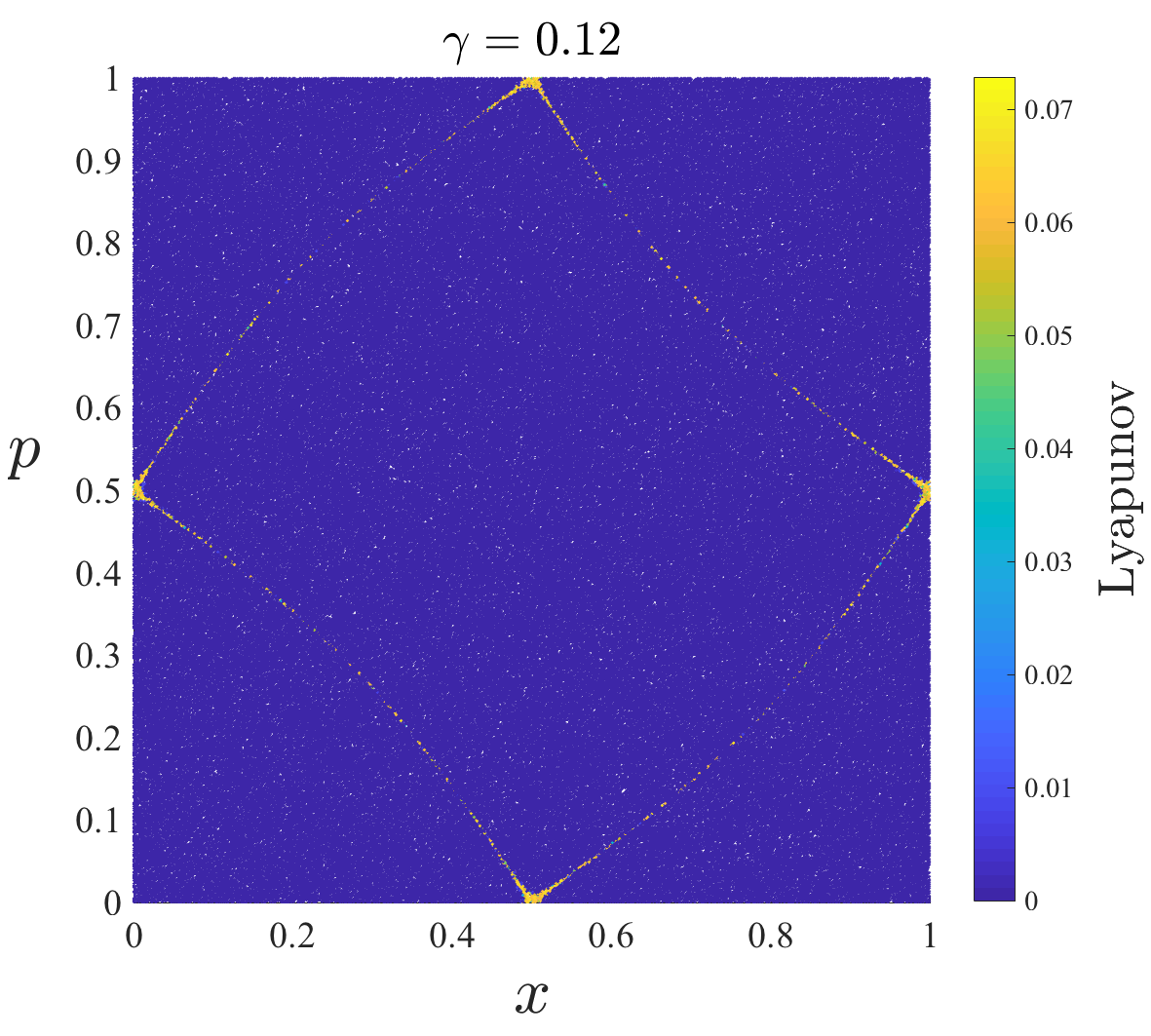}
\includegraphics[height=4.8cm,angle=0]{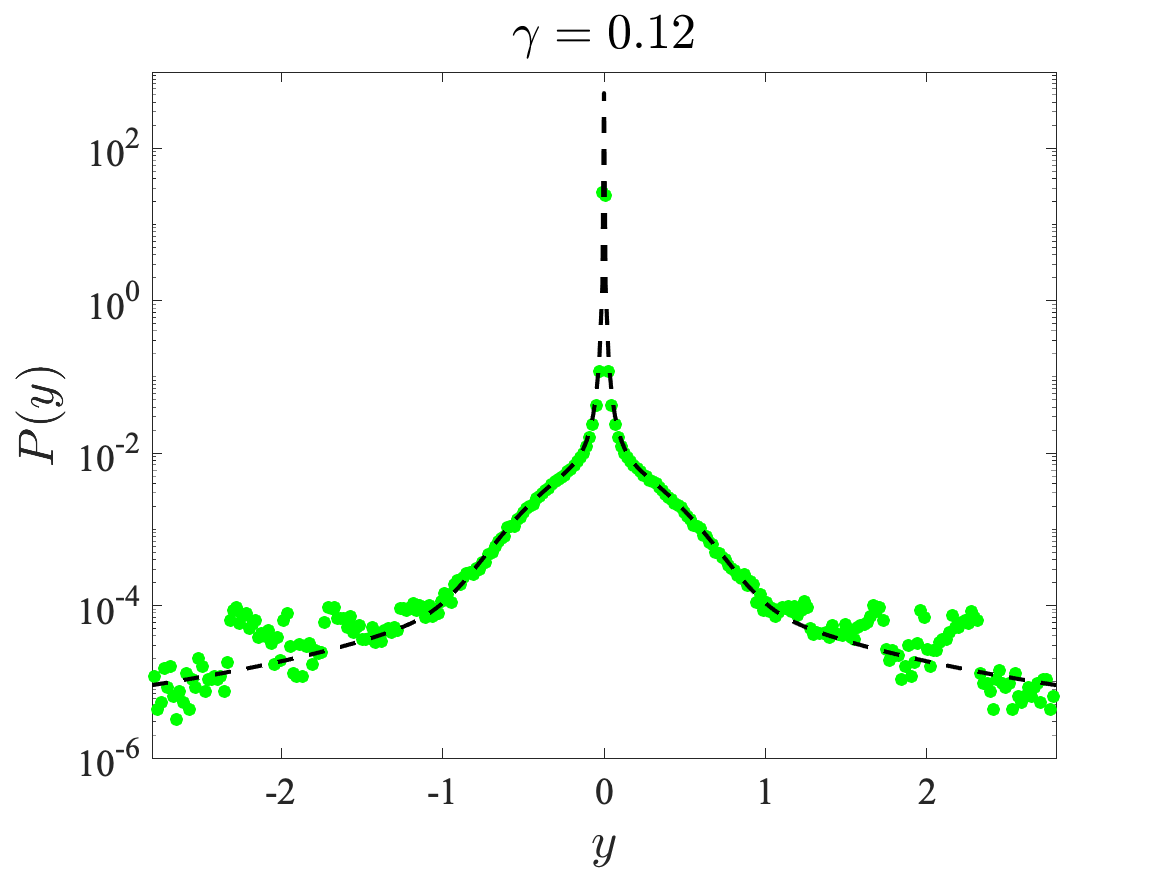}\\
\includegraphics[height=5cm,angle=0]{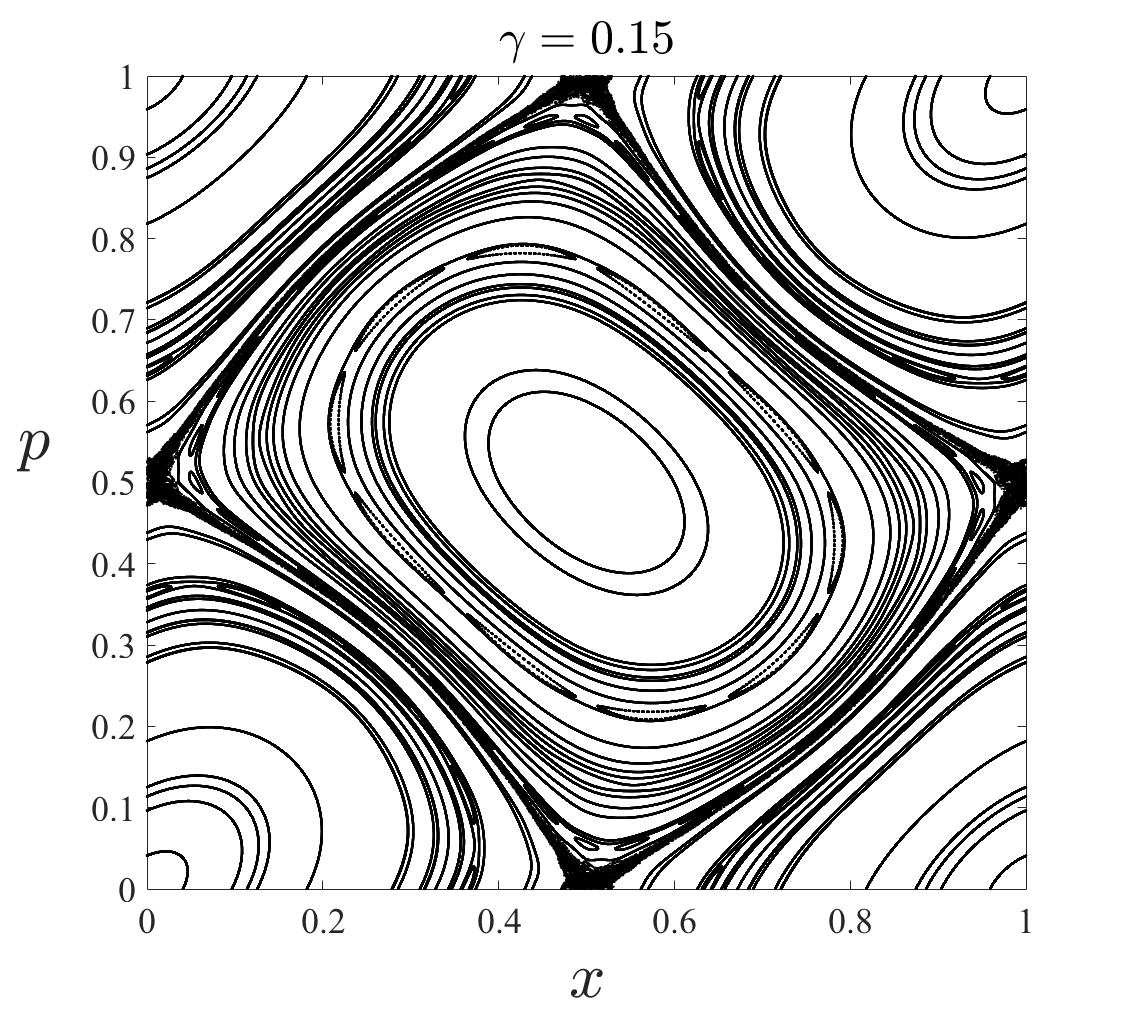}
\includegraphics[height=5cm,angle=0]{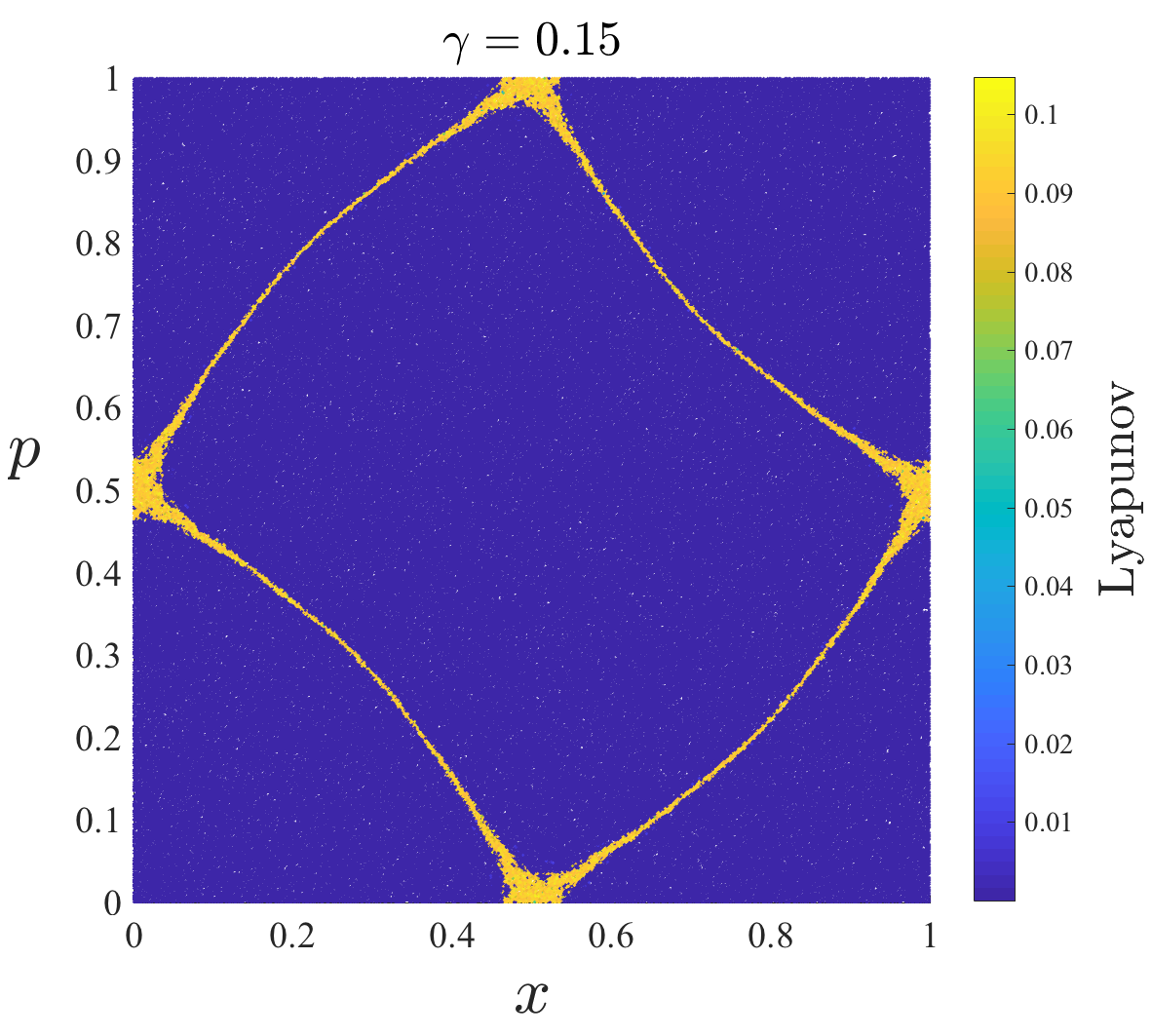}
\includegraphics[height=4.8cm,angle=0]{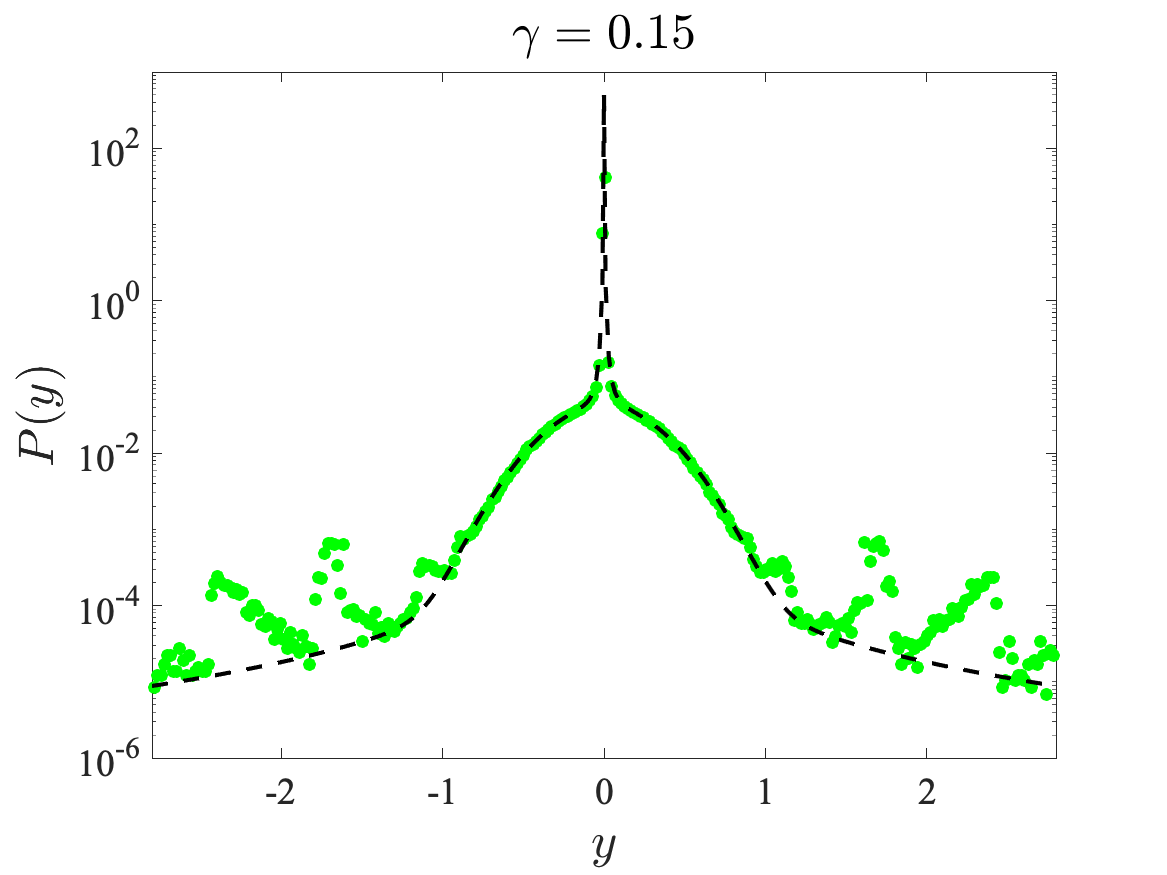}\\
\includegraphics[height=5cm,angle=0]{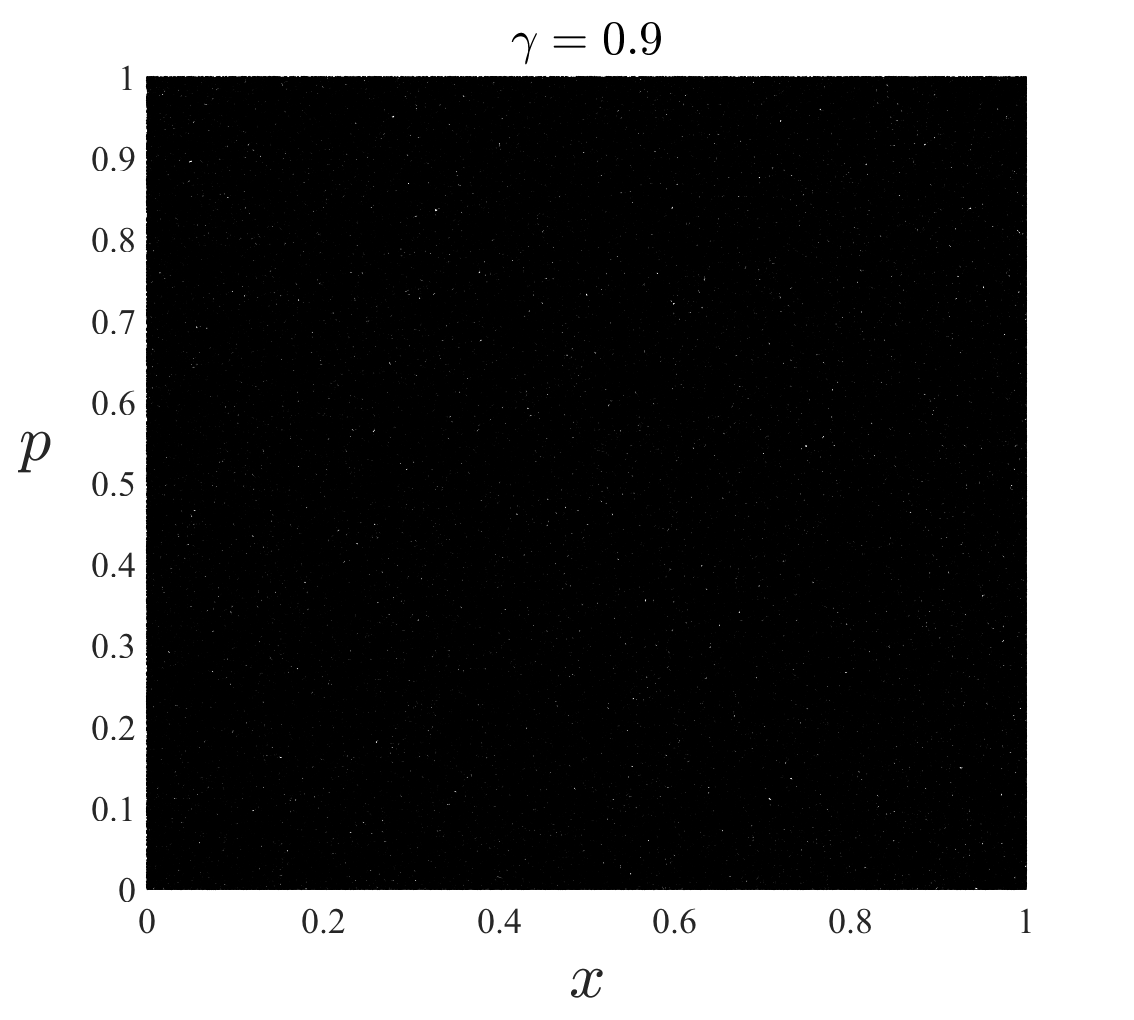}
\includegraphics[height=5cm,angle=0]{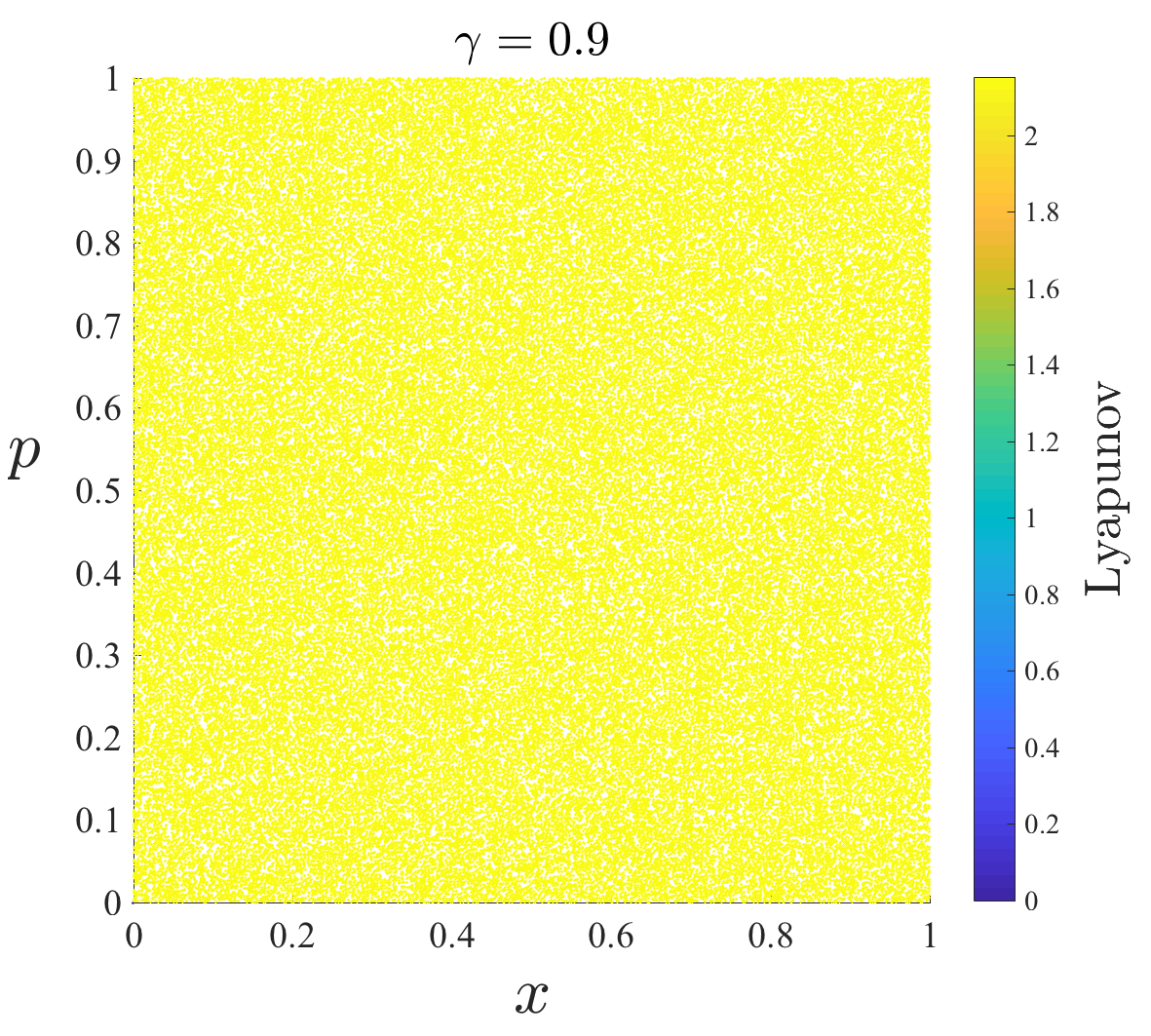}
\includegraphics[height=4.8cm,angle=0]{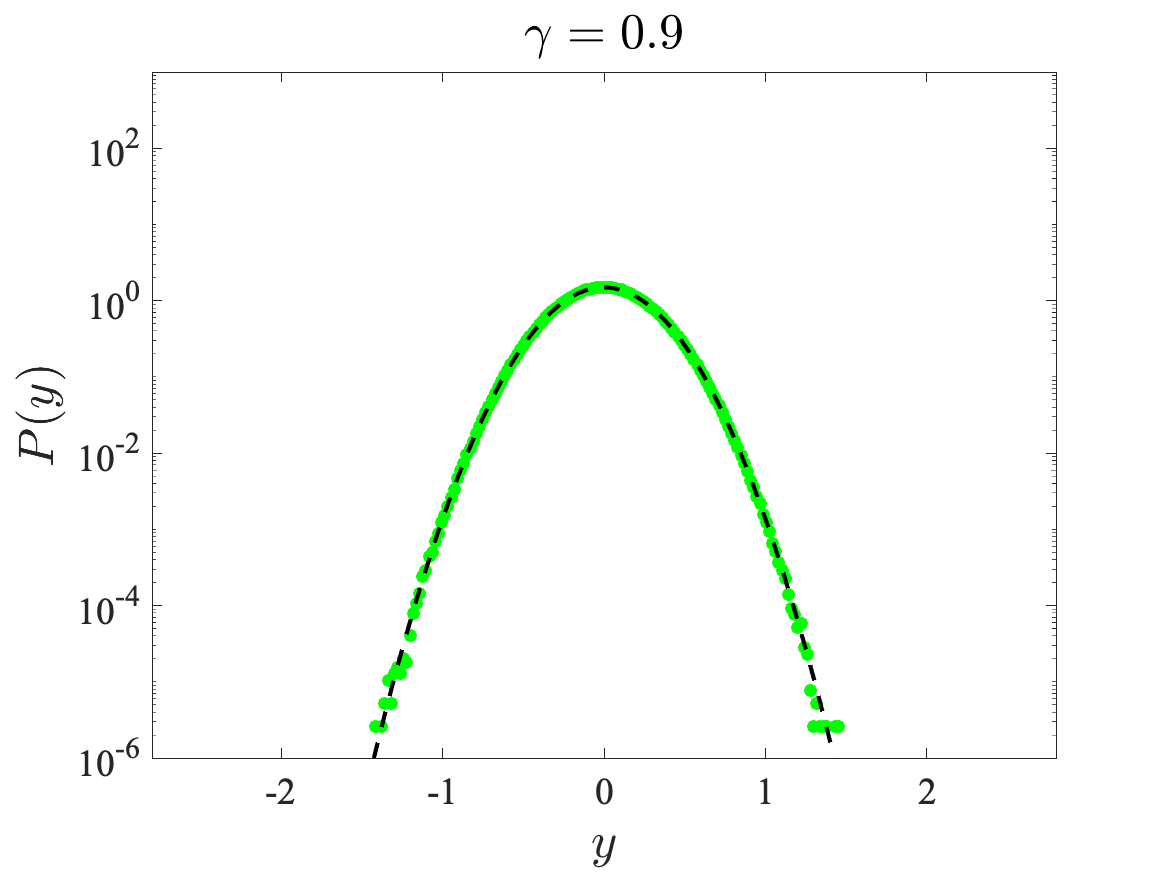}\\
\caption{\label{fig:Fig1} {\small (Color online) 
{\it Left column:}~Phase space portrait of the Harper map for various $\gamma$ values;  
40-50 initial conditions have been used in each case. 
{\it Middle column:}~Lyapunov diagrams for the same values of $\gamma$. 
The Lyapunov exponents have been calculated over 500000 time steps using 500000 initial 
conditions taken randomly from the entire phase space. 
{\it Right column:}~Probability distributions obtained from the same values of $\gamma$. 
In all cases, the number of initial conditions is larger than $3\times 10^{7}$ in order to achieve 
good statistics.}}
\end{figure}

As discussed in \cite{harper}, in the Harper map phase-space plots, a separatrix defined by $H_0(p,x)=0$ 
appears and forms a square symmetry. When $\gamma >0$, this separatrix is destroyed and 
becomes a mesh of finite thickness inside which the dynamics is chaotic. This region grows 
as $\gamma$ increases, making the region of the regular motion to shrink. This behavior is evident 
from the first column of Fig.~\ref{fig:Fig1} for some representative $\gamma$ values.  
Then, in the second column, one can see the Lyapunov diagram for the same values of $\gamma$. 
The genesis and increasing domination of the chaotic sea can be clearly seen as $\gamma$ 
increases. Finally, the last column exhibits the corresponding probability distributions. 
As discussed extensively in \cite{Ruiz,Ruiz2017b}, these distributions can be well approximated by  
a linear combination of one Gaussian and one $q$-Gaussian distributions, namely, 

\begin{equation}
P(y)=\alpha_{q_{1}}P_{q_{1}}(y;\mu_{q_{1}},\sigma_{q_{1}})+\alpha_{q_{2}}P_{q_{2}}(y;\mu_{q_{2}},
\sigma_{q_{2}})
\label{Pq}
\end{equation}
where $q_{1}=1.935\pm0.005$ and $q_{2}=1$. 
In Eq.~\ref{Pq}, the contribution ratios $\alpha_{q_{1}}$ and $\alpha_{q_{2}}$ can be evaluated from the 
phase-space occupation ratios of initial conditions located in the stability islands and chaotic sea 
detected from the Lyapunov color map, respectively. Therefore, these parameters are not fitting 
parameters, but determined directly from the dynamics of the system. The $q$-Gaussian distribution 
with $q= 1.935\pm0.005$ originates from the initial conditions of the stability islands, and initial 
conditions from the chaotic sea contributes to the Gaussian distribution. The obtained results for 
the parameters are given in Table.~\ref{Table1}. These results clearly show that the Harper is in the 
same universality class of the standard map and the web map, and therefore provide one more argument 
pointing to the robustness of the $q$-Gaussian with $q= 1.935\pm0.005$.

\begin{table}[H]
  \caption{Obtained results for the Harper map for four representative values of $\gamma$.}
\centering
\begin{tabular}{c|c|c||c|c|}
 \cline{2-5}
     \cline{2-5}
  & $\gamma=0.05$ & $\gamma=0.12$ & $\gamma=0.15$ & $\gamma=0.9$  \\
\cline{2-5}
\cline{2-5}
\hline
\hline
\multicolumn{1}{|c||}{$q_1$}& 1.935&  1.935& 1.935& 1.935 \\
\hline
\multicolumn{1}{|c||}{$q_2$}& 1&  1& 1 & 1   \\
\hline
\multicolumn{1}{|c||}{$\alpha_{q_1}$}& 1 & 0.9960  & 0.9713 & 0  \\
\hline
\multicolumn{1}{|c||}{$\alpha_{q_2}$}& 0  & 0.0040 & 0.0287 & 1  \\
\hline
\multicolumn{1}{|c||}{$B_{q_1}$}& $65\times 10^5$& $25\times 10^5$  & $25\times 10^5$ & -   \\
\hline
\multicolumn{1}{|c||}{$B_{q_2}$}& - &  $5.3$& $5.7$ & $7.0$  \\
\hline
\end{tabular}
\label{Table1}
\end{table}

\begin{figure}[H]
\centering
\includegraphics[height=4.87cm,angle=0]{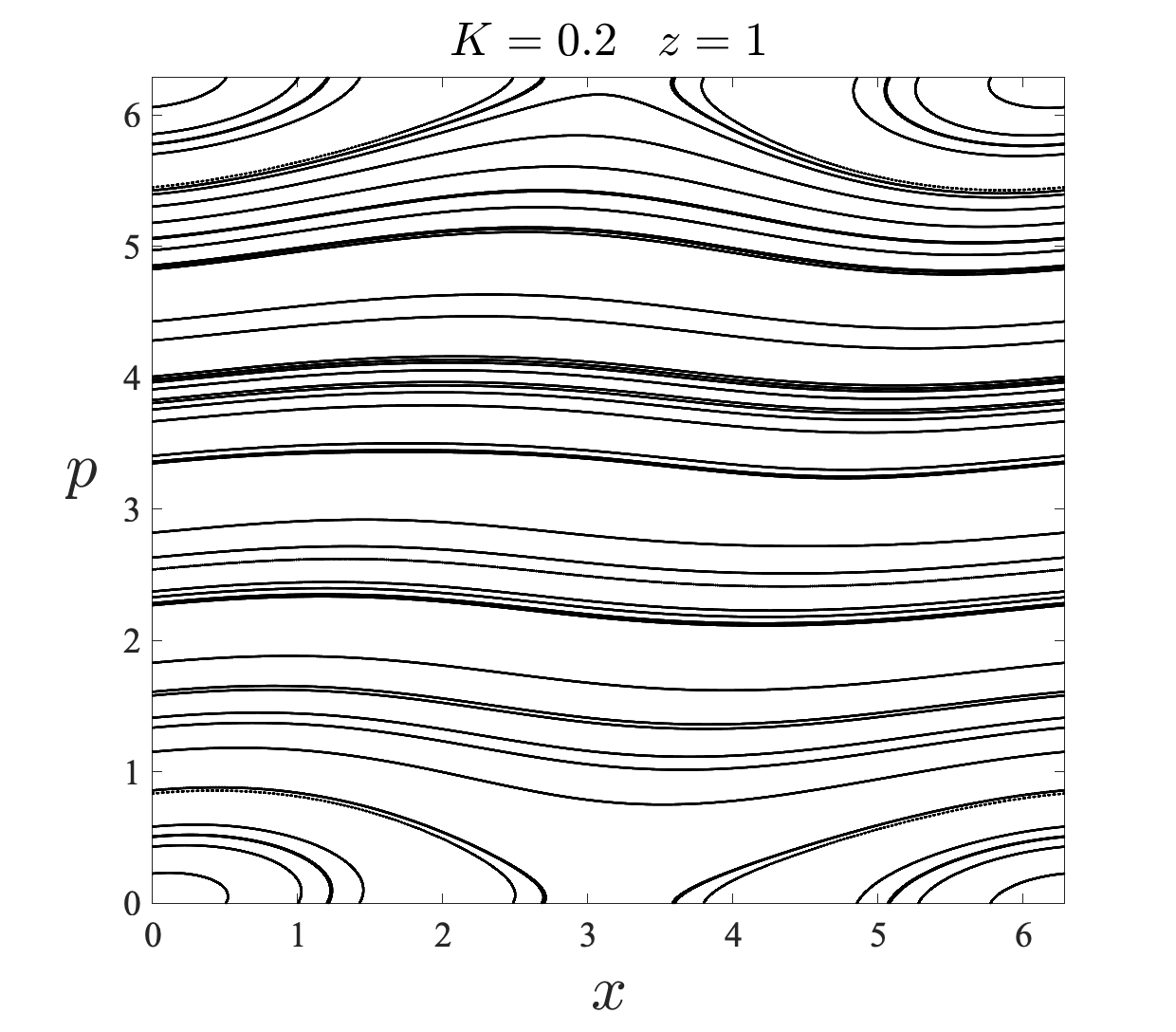}
\includegraphics[height=4.87cm,angle=0]{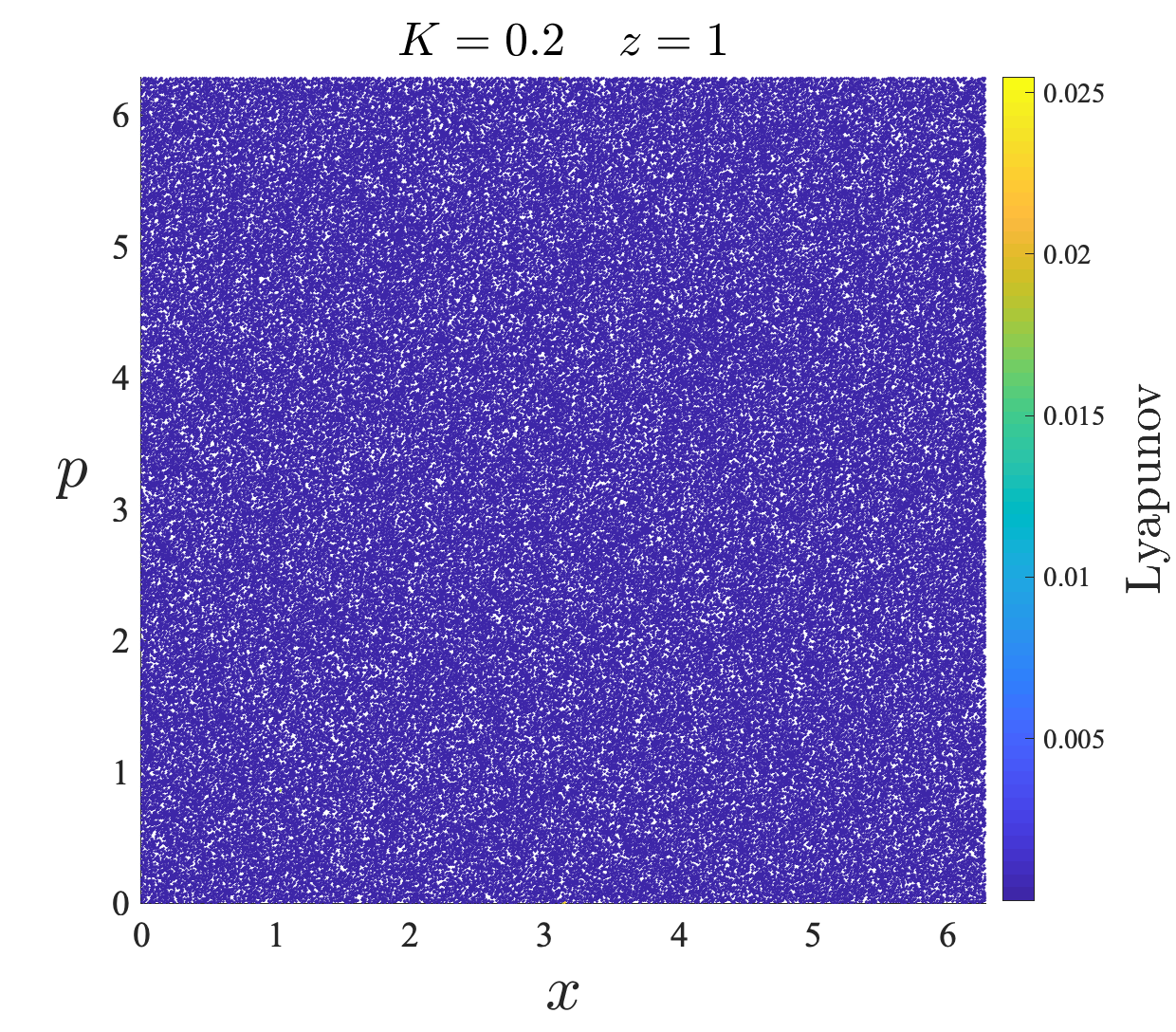}
\includegraphics[height=4.8cm,angle=0]{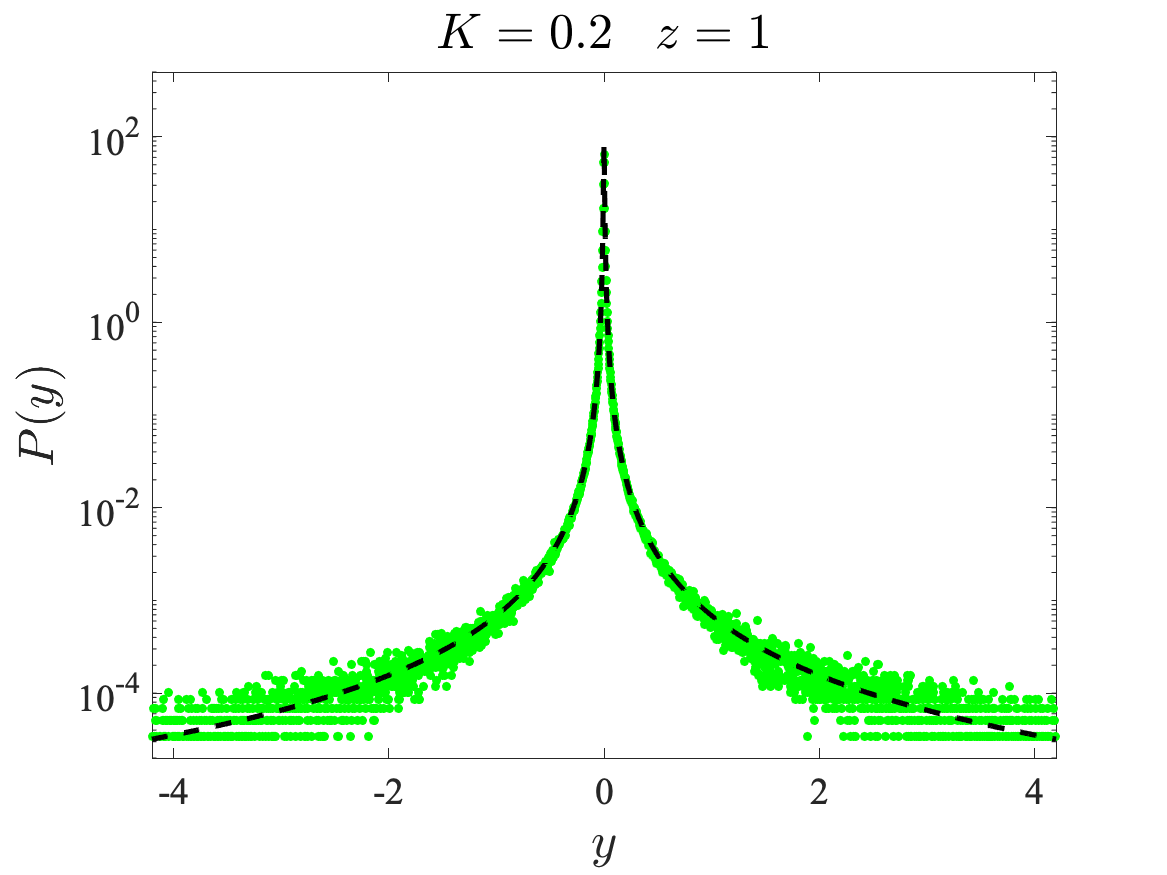}\\
\includegraphics[height=4.87cm,angle=0]{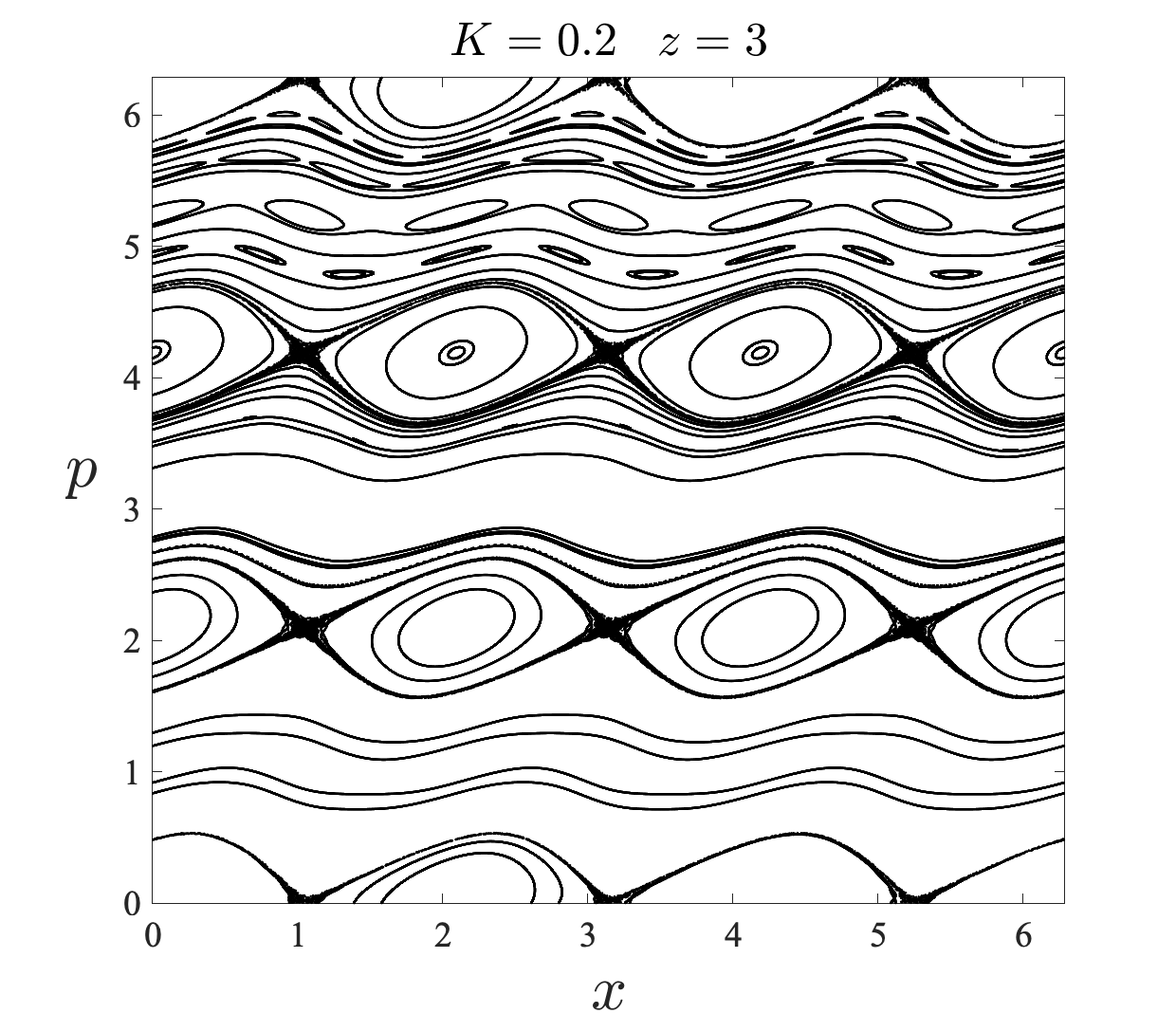}
\includegraphics[height=4.87cm,angle=0]{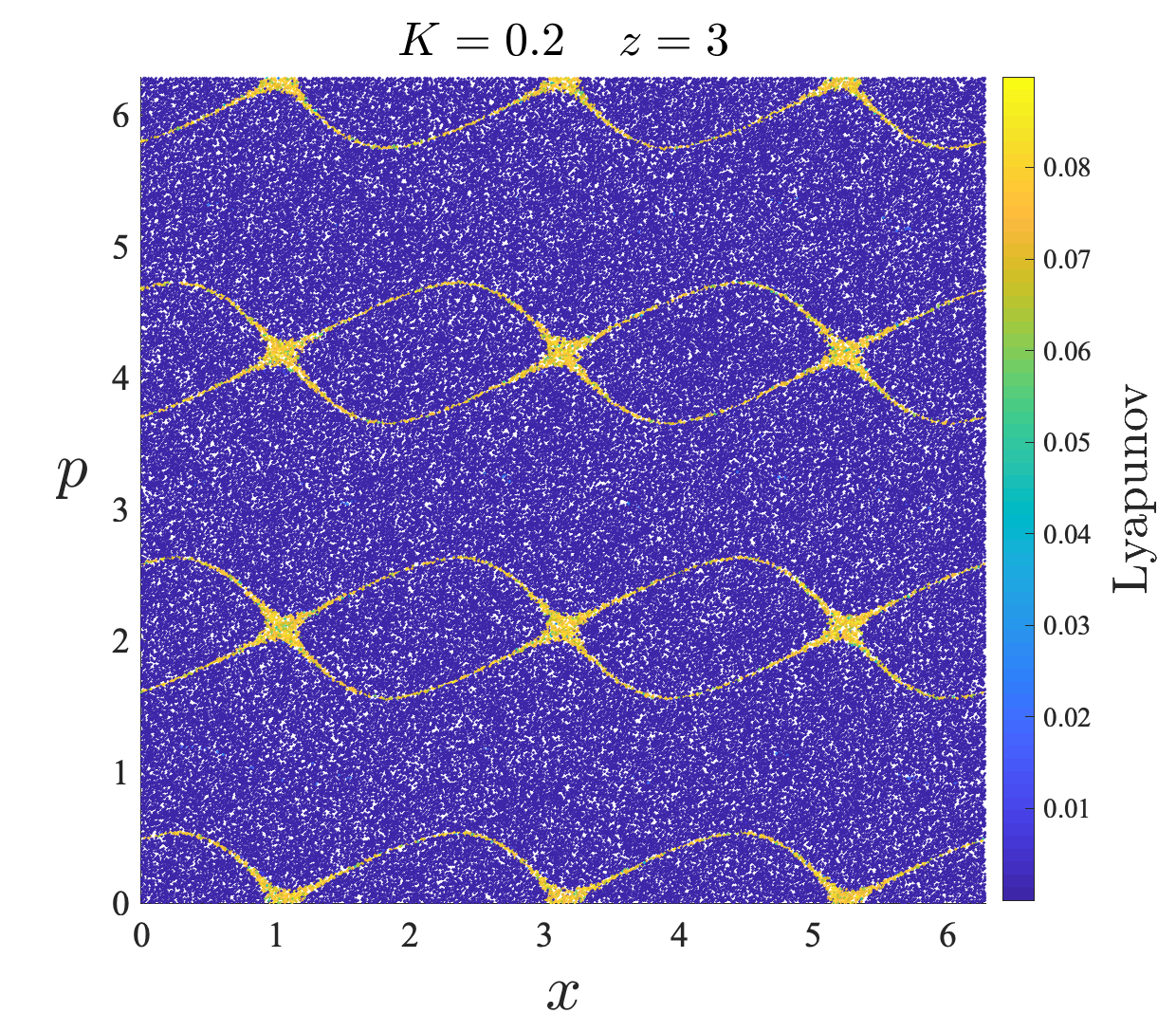}
\includegraphics[height=4.8cm,angle=0]{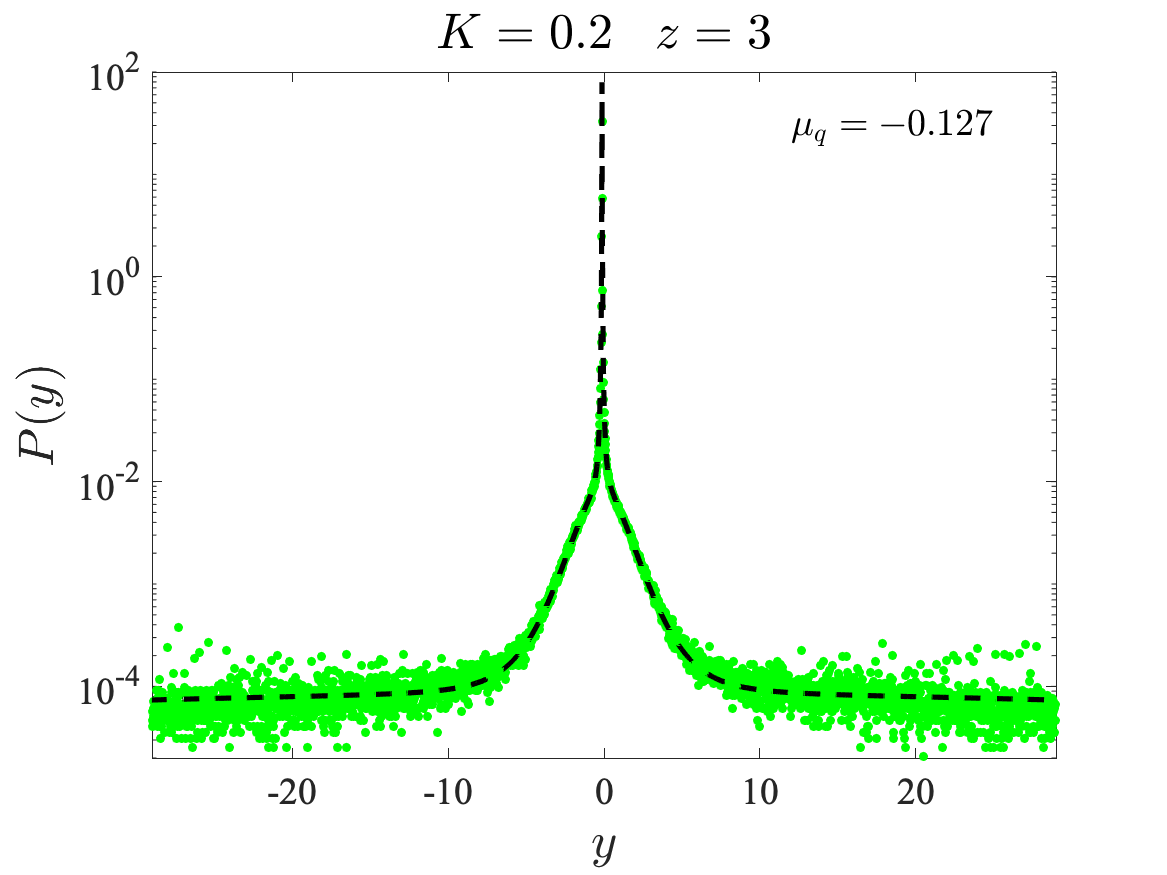}\\
\includegraphics[height=4.87cm,angle=0]{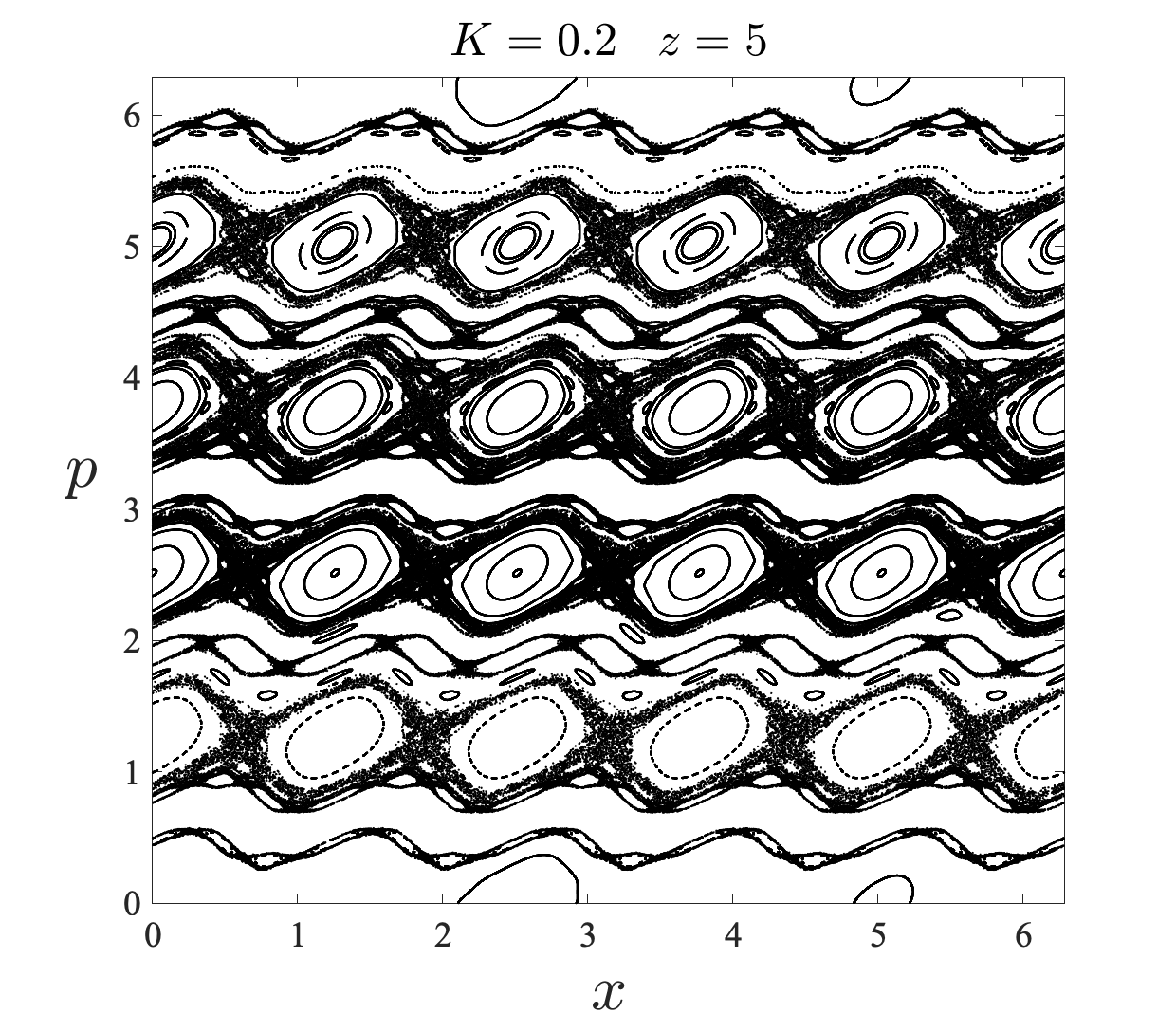}
\includegraphics[height=4.87cm,angle=0]{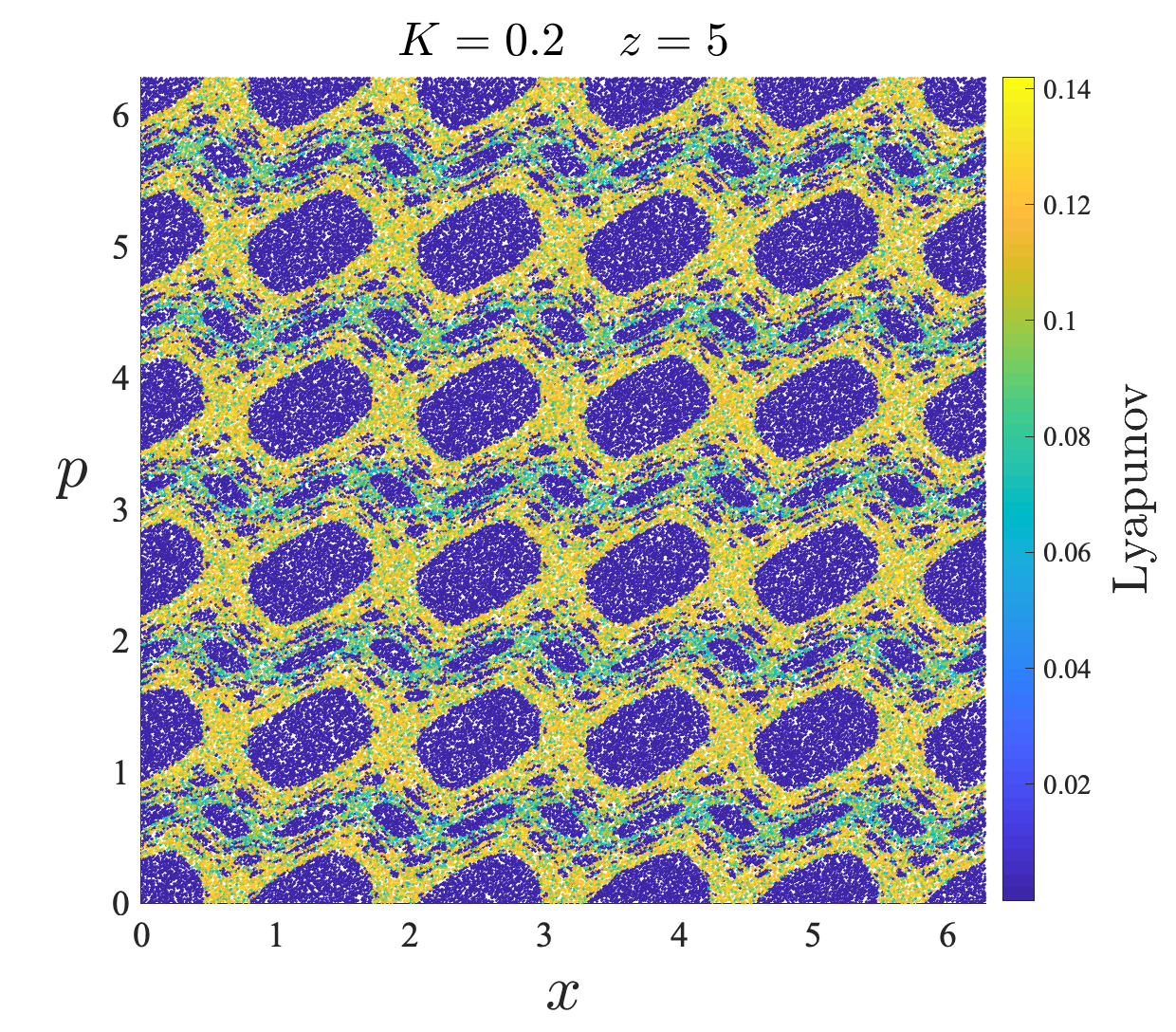}
\includegraphics[height=4.8cm,angle=0]{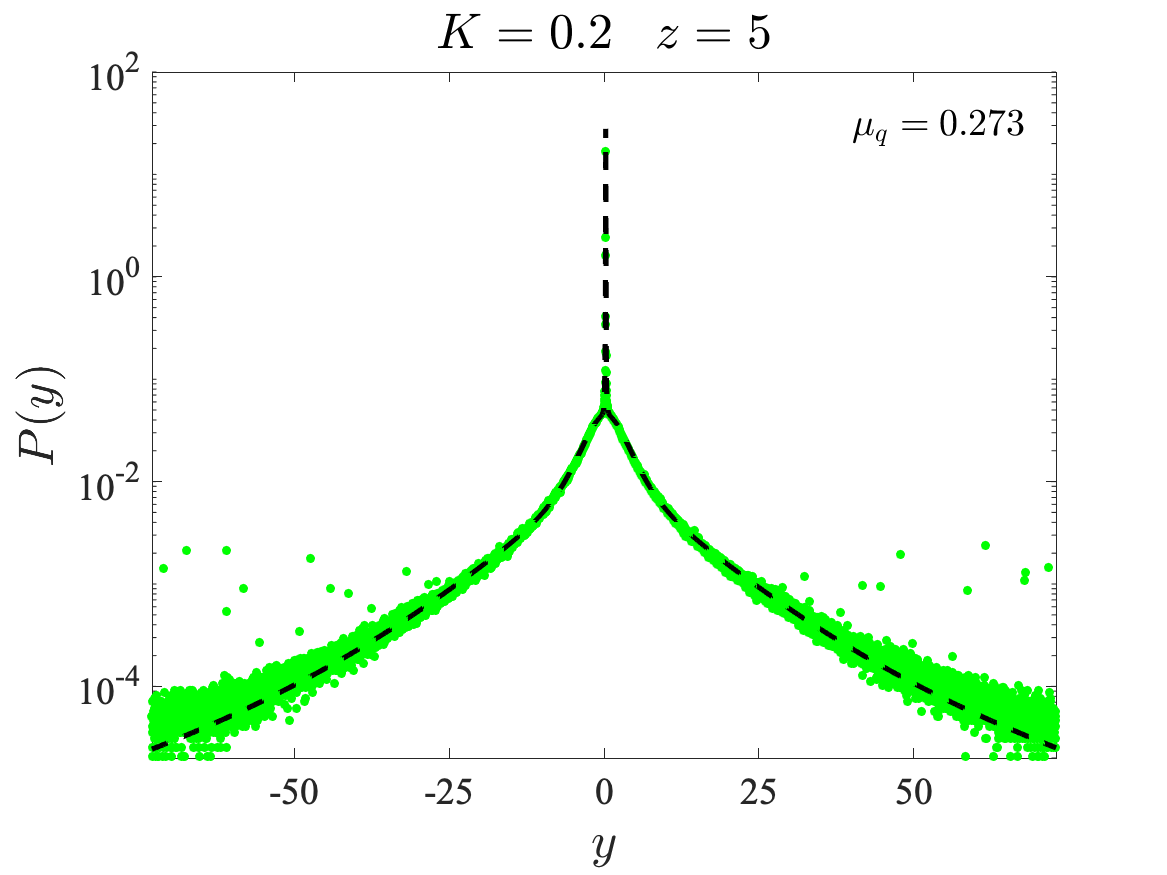}\\
\includegraphics[height=4.87cm,angle=0]{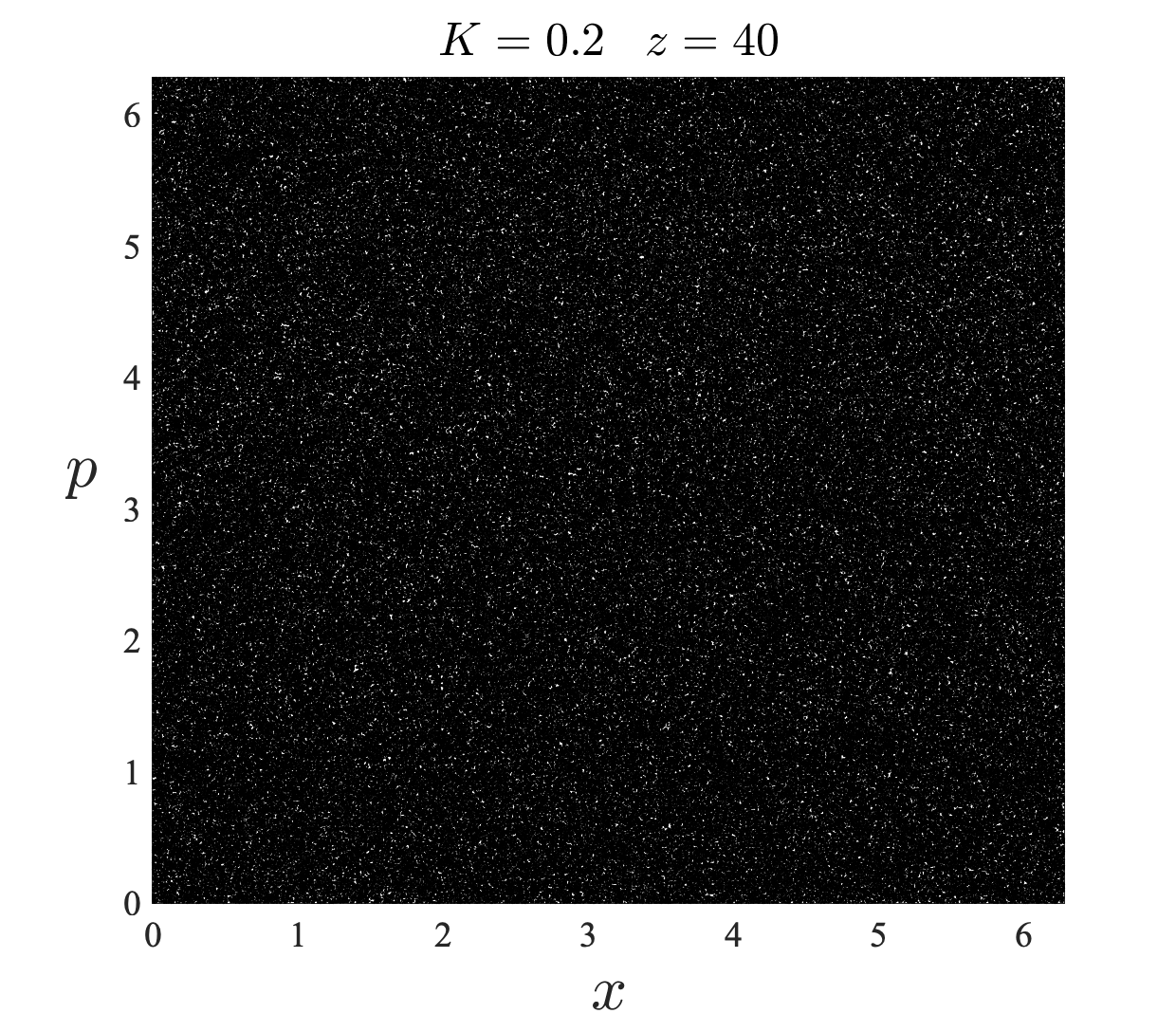}
\includegraphics[height=4.87cm,angle=0]{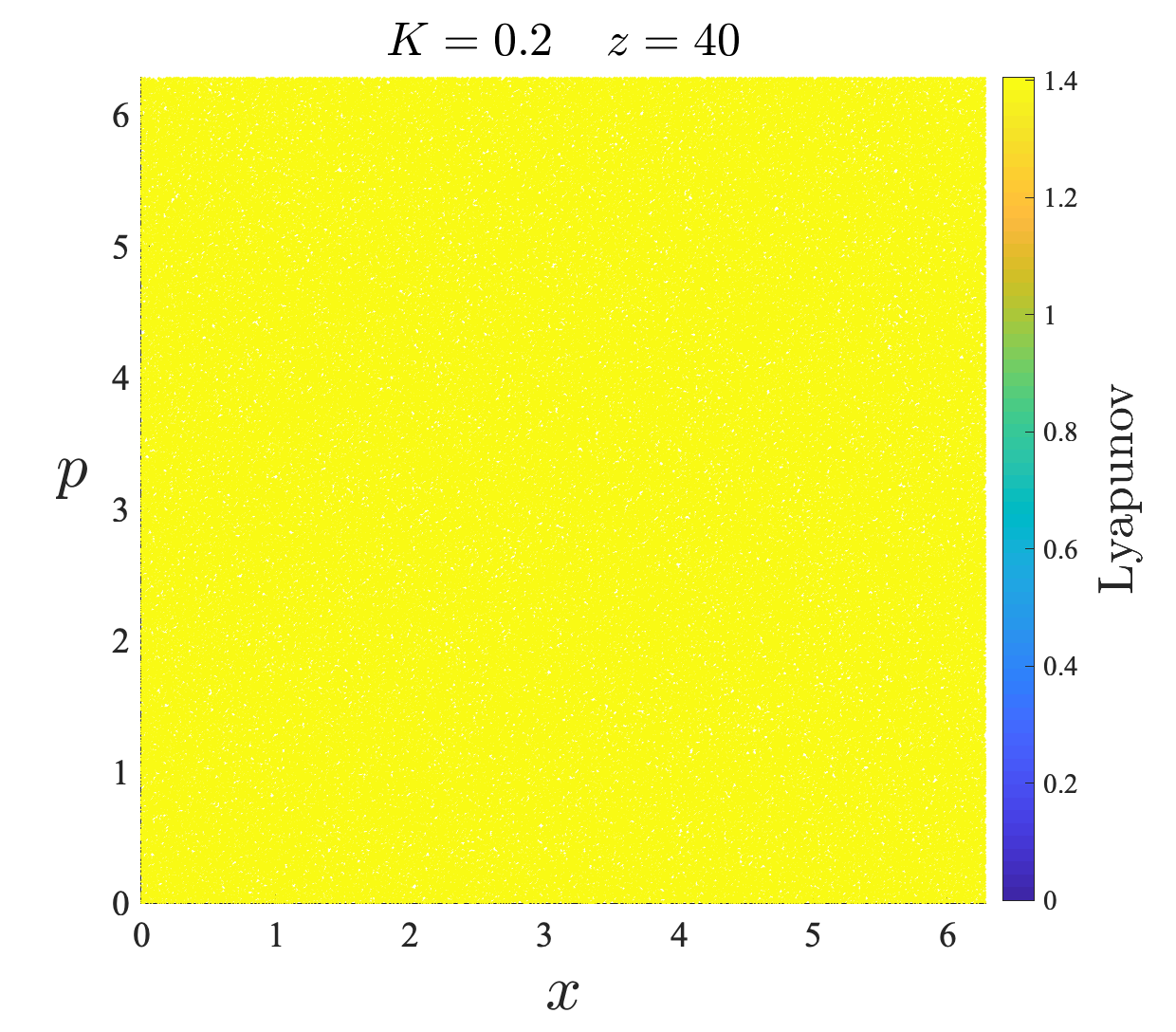}
\includegraphics[height=4.8cm,angle=0]{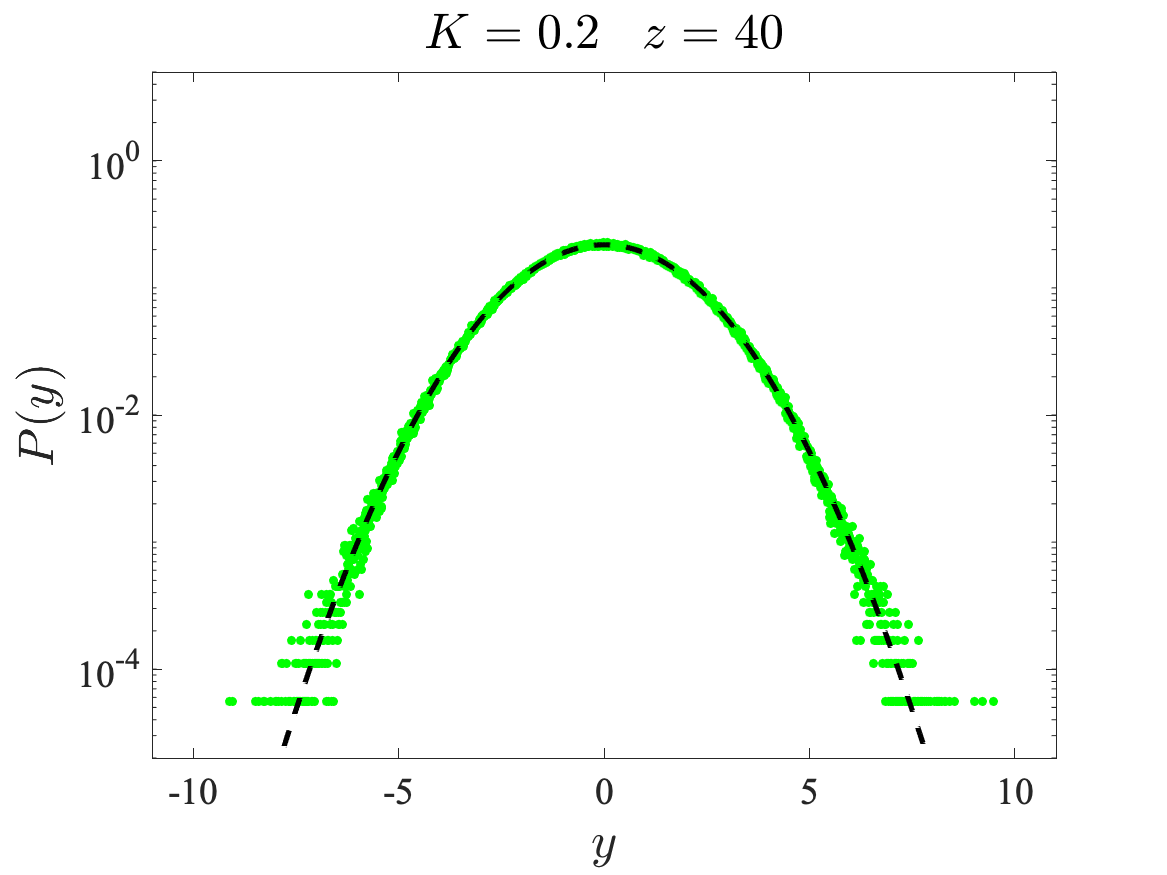}\\
\caption{\label{fig:Fig2} {\small (Color online) 
{\it Left column:}~Phase space portrait of the $z$-standard map for $K=0.2$ with various $z$ values; 
40-50 initial conditions have been used in each case. 
{\it Middle column:}~ Lyapunov diagrams for the same cases. 
The Lyapunov exponents have been calculated over 500000 time steps using 500000 initial 
conditions taken randomly from the entire phase space. 
{\it Right column:}~Probability distributions obtained for the same cases. 
In all cases, the number of initial conditions is larger than $3\times 10^{7}$ in order to achieve 
good statistics. }}
\label{Figure2}
\end{figure}

\begin{figure}[H]
\centering
\includegraphics[height=4.87cm,angle=0]{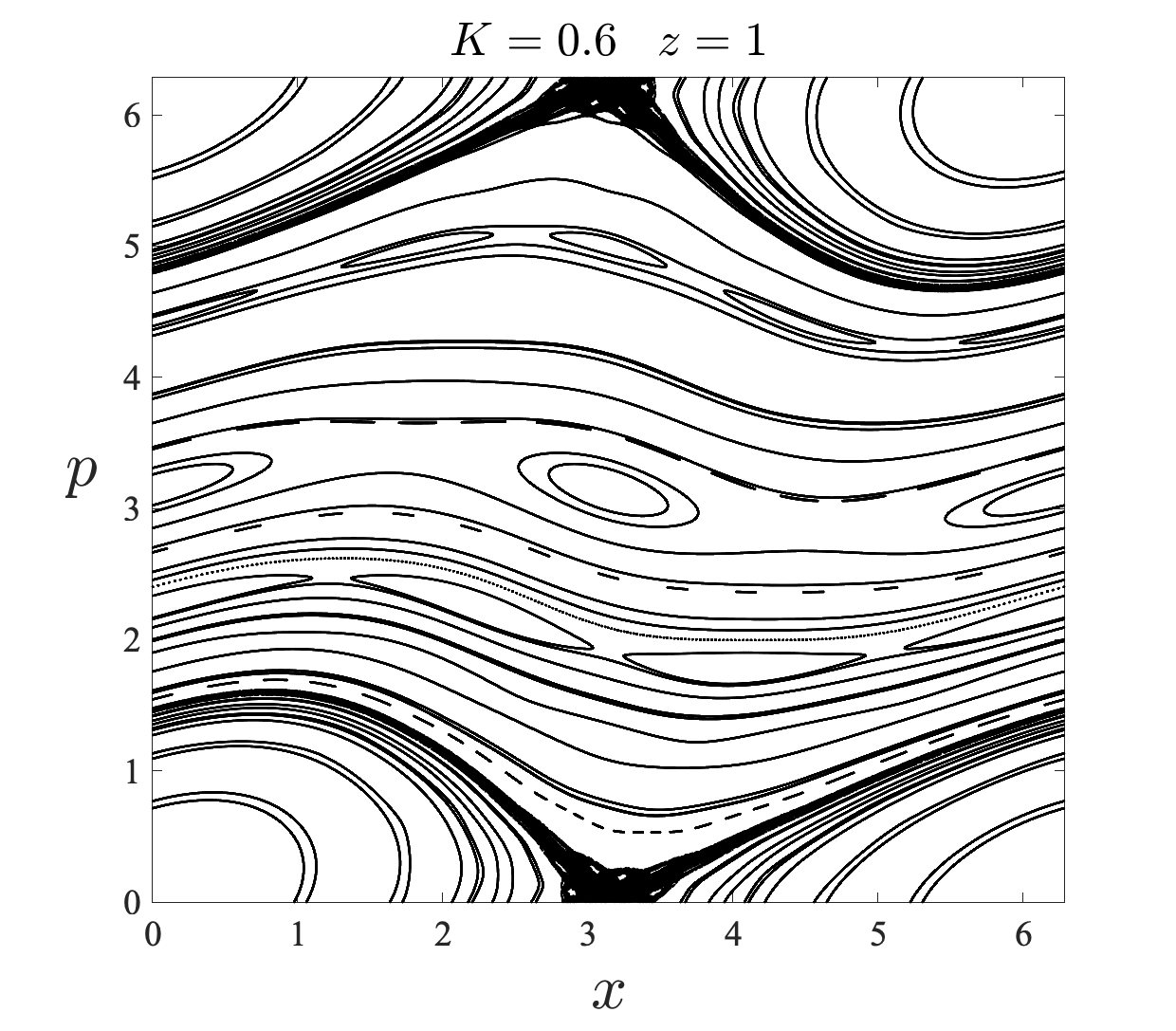}
\includegraphics[height=4.87cm,angle=0]{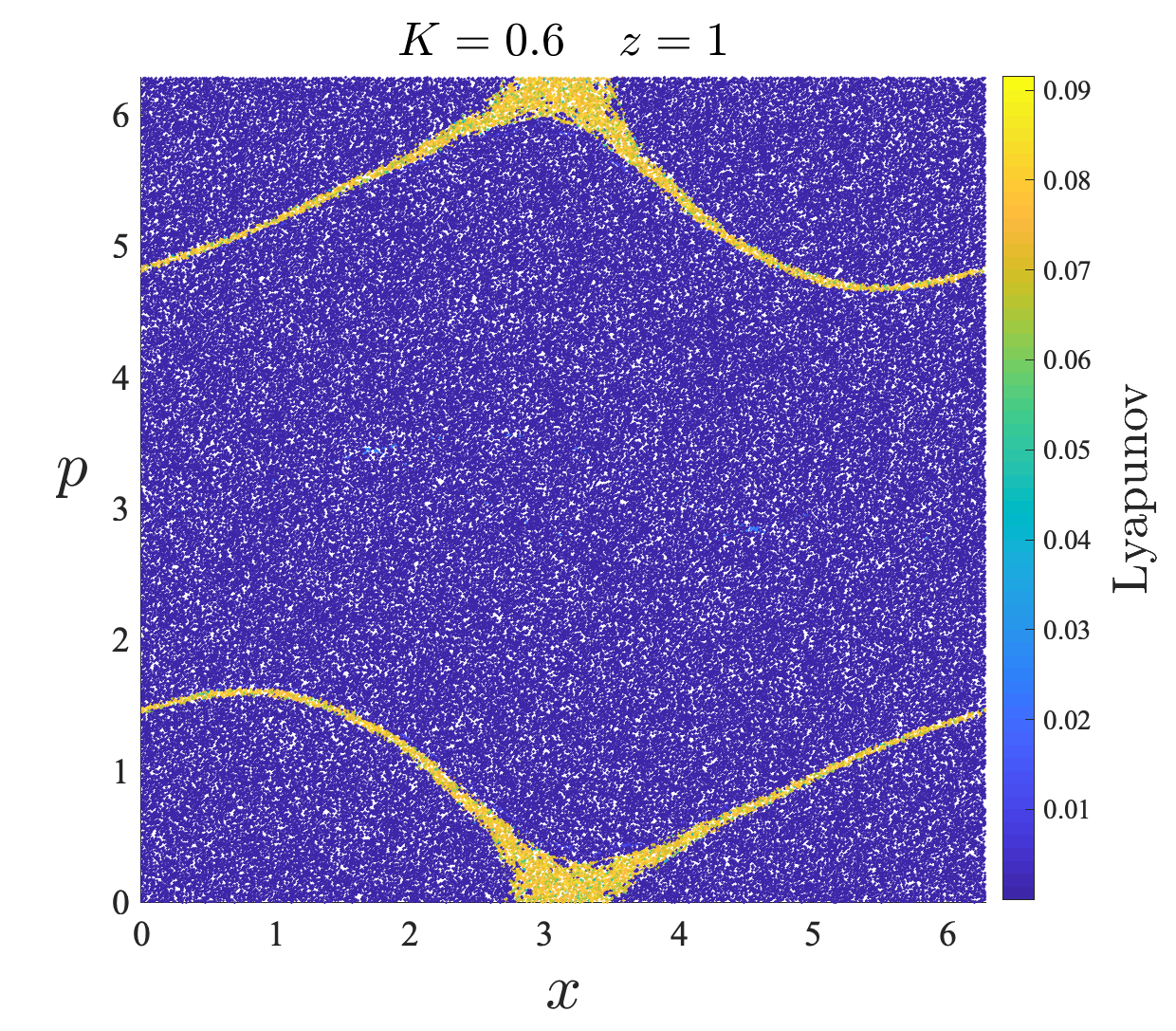}
\includegraphics[height=4.8cm,angle=0]{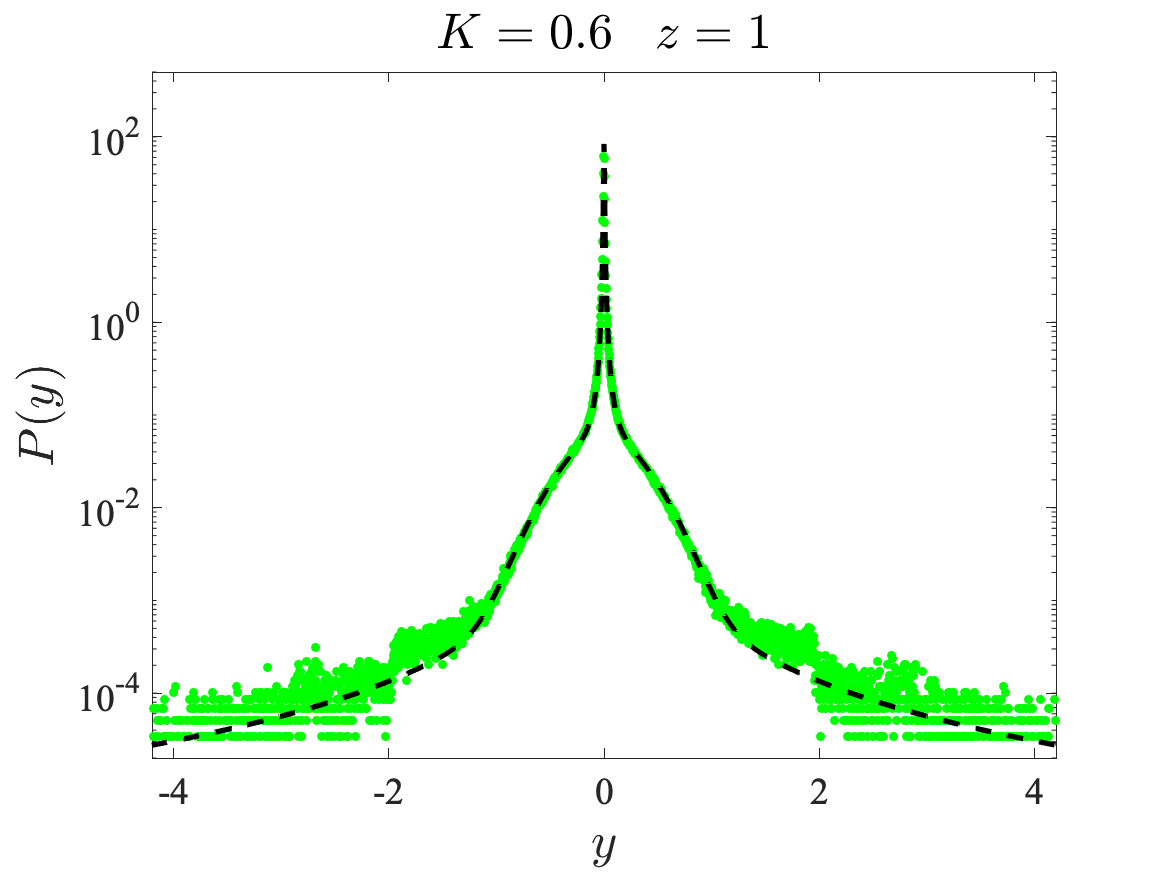}\\
\includegraphics[height=4.87cm,angle=0]{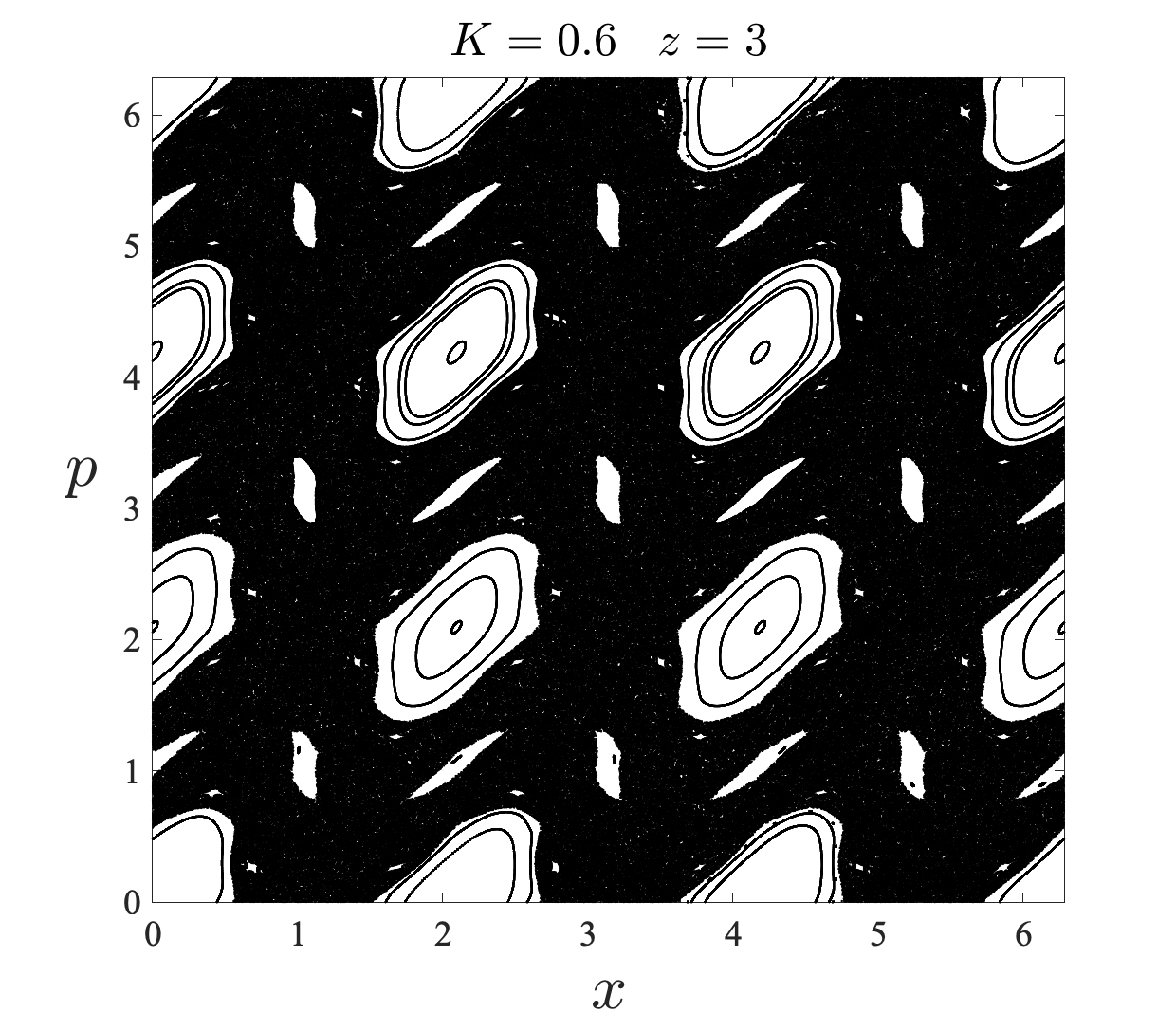}
\includegraphics[height=4.87cm,angle=0]{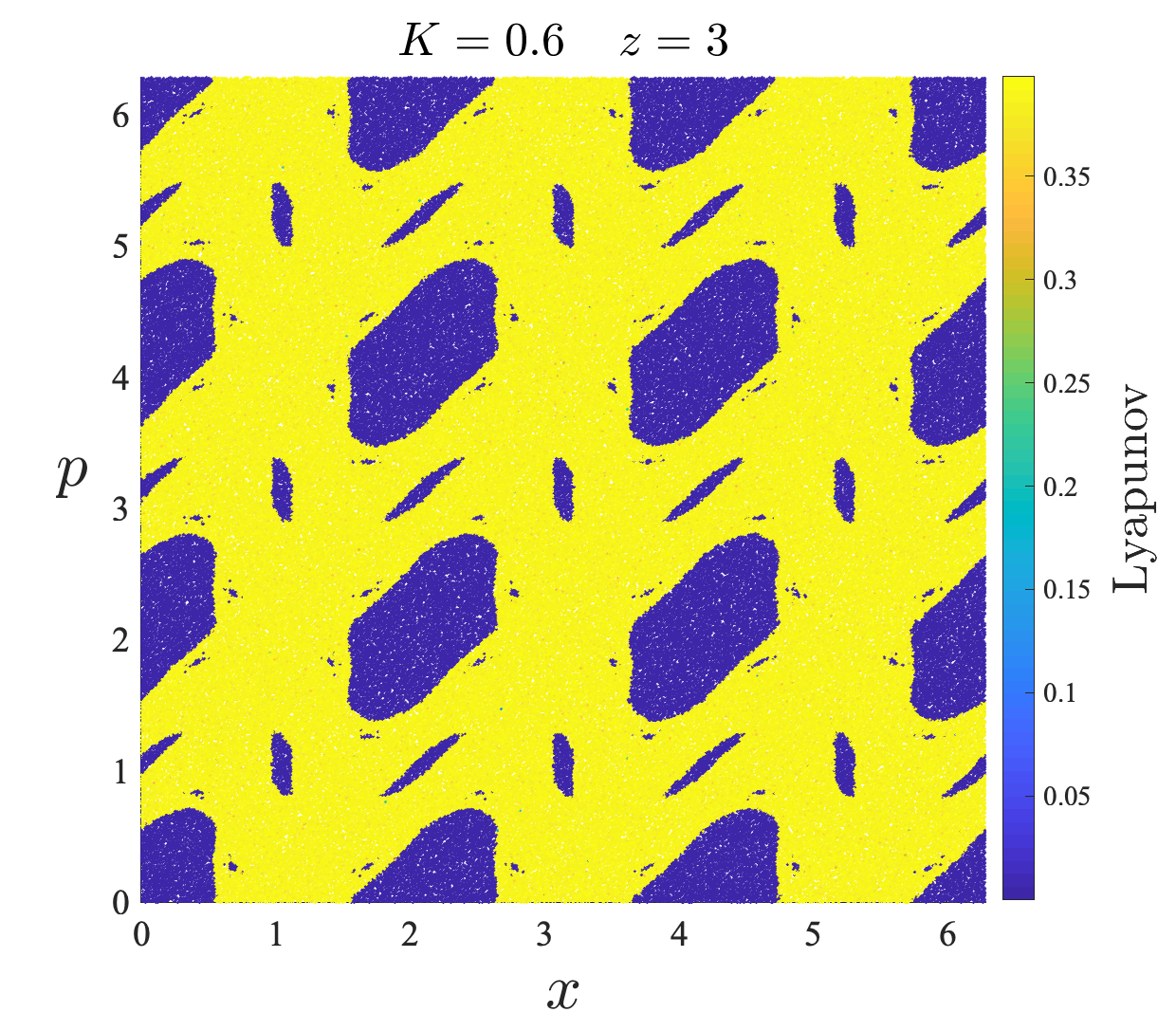}
\includegraphics[height=4.8cm,angle=0]{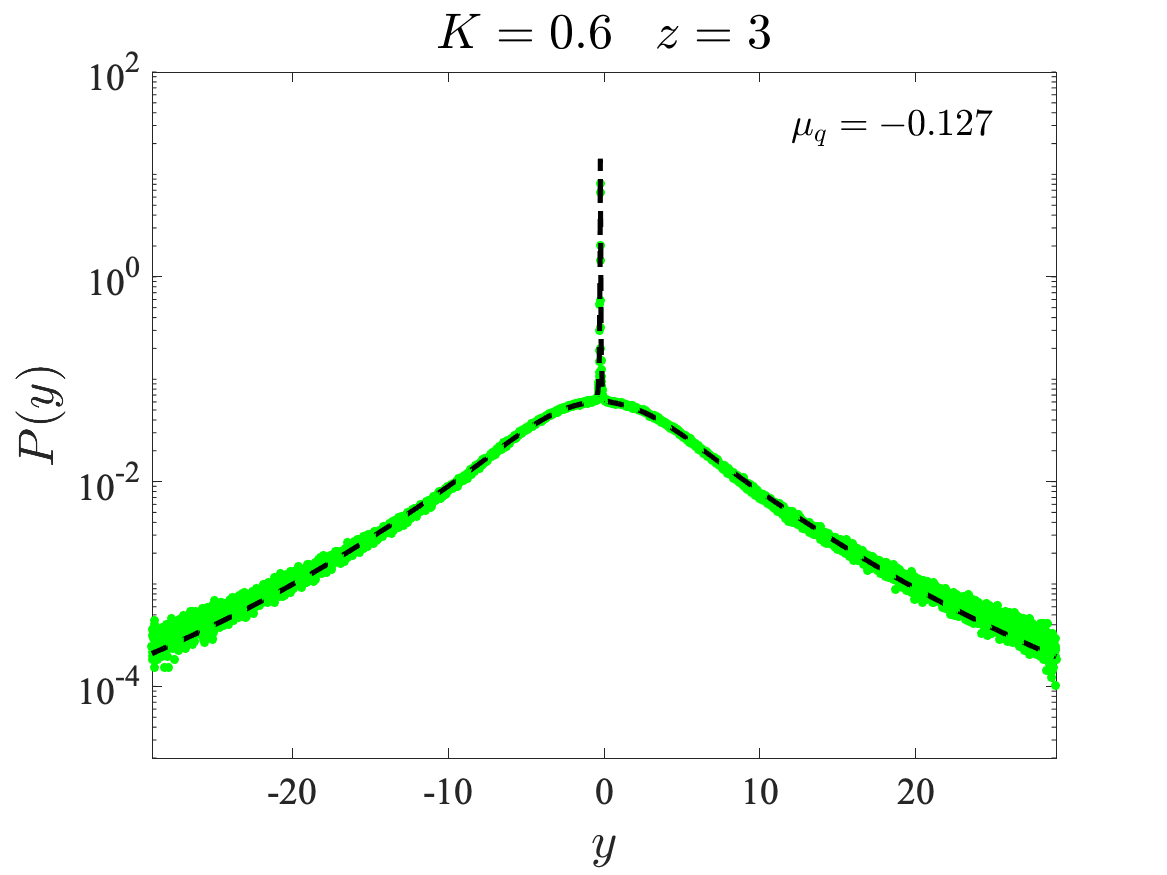}\\
\includegraphics[height=4.87cm,angle=0]{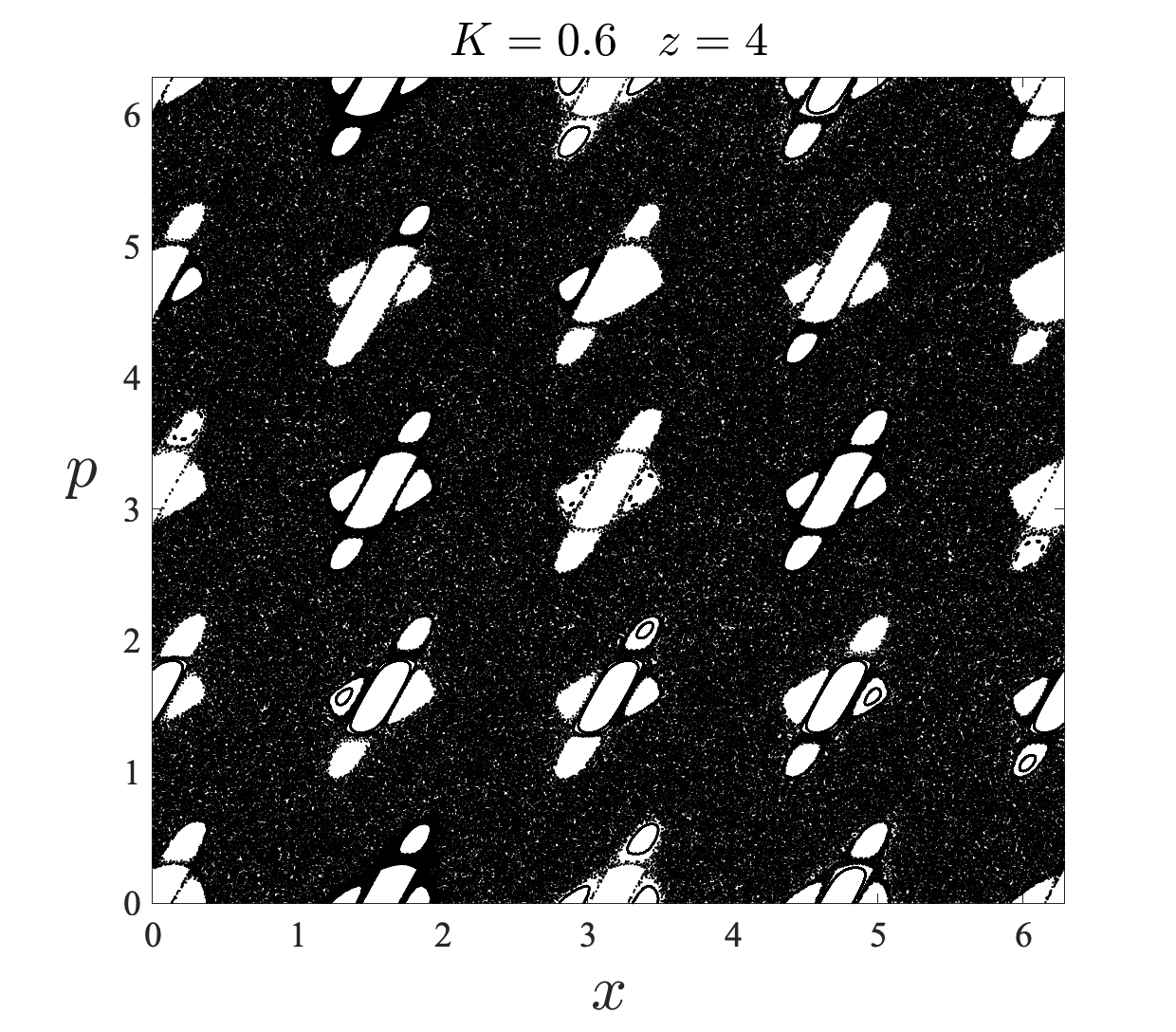}
\includegraphics[height=4.87cm,angle=0]{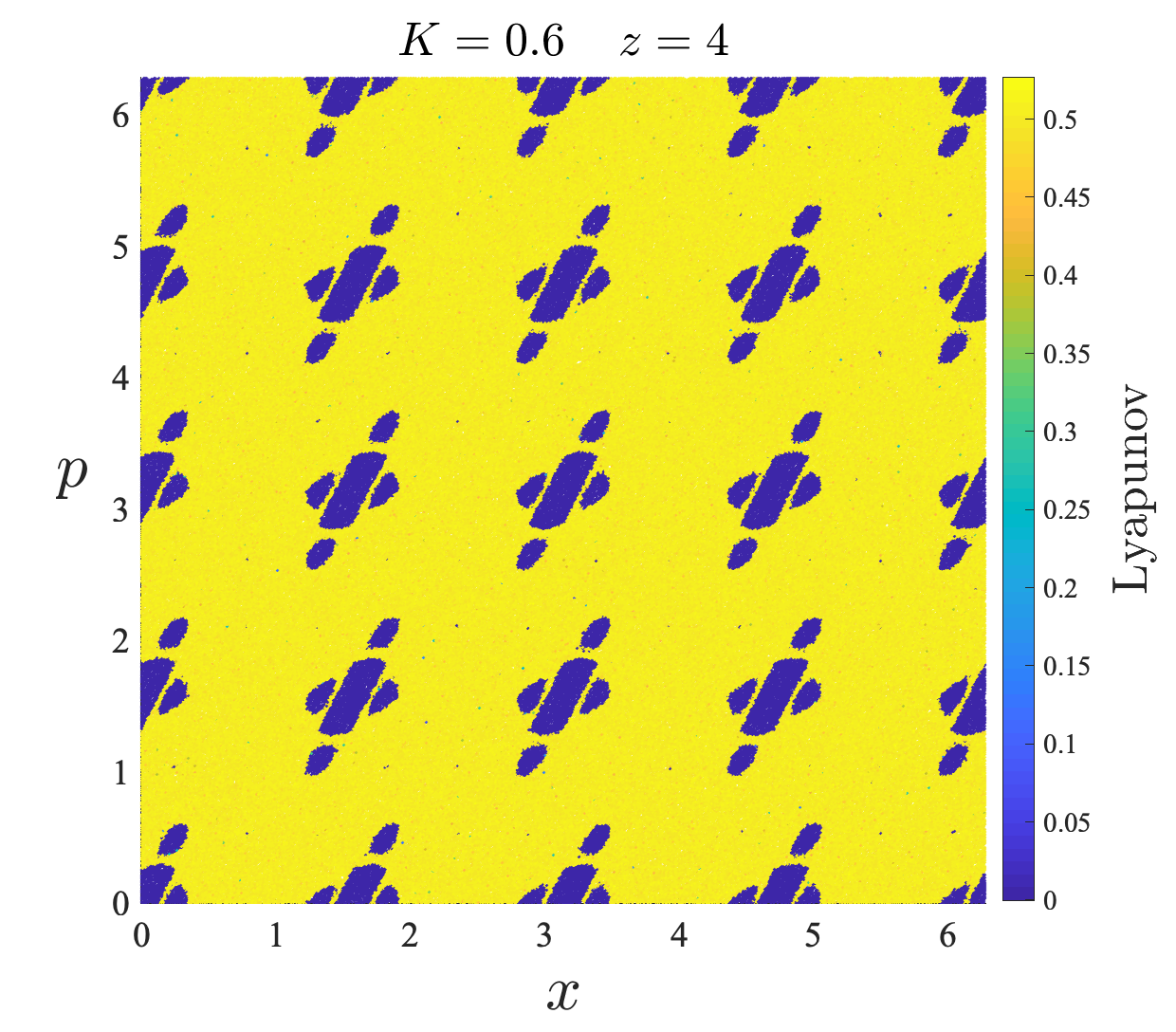}
\includegraphics[height=4.8cm,angle=0]{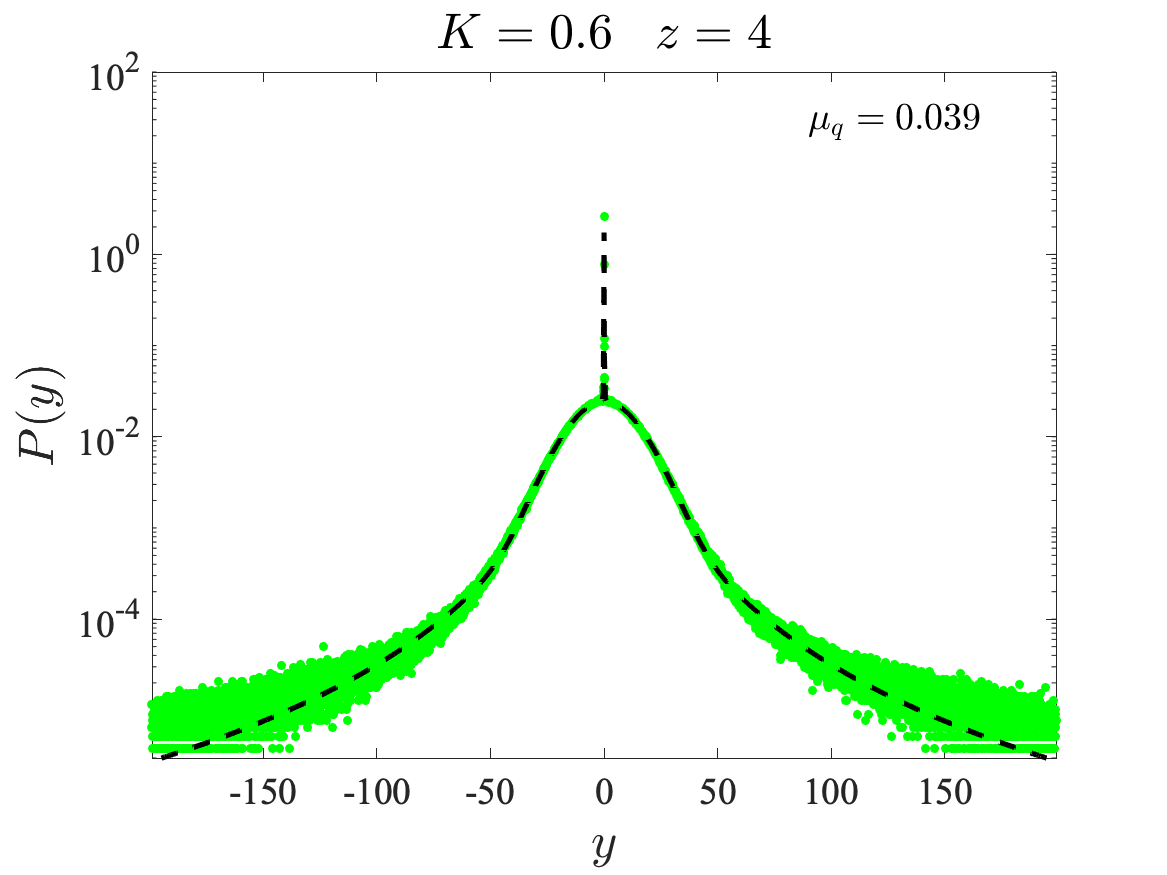}\\
\includegraphics[height=4.87cm,angle=0]{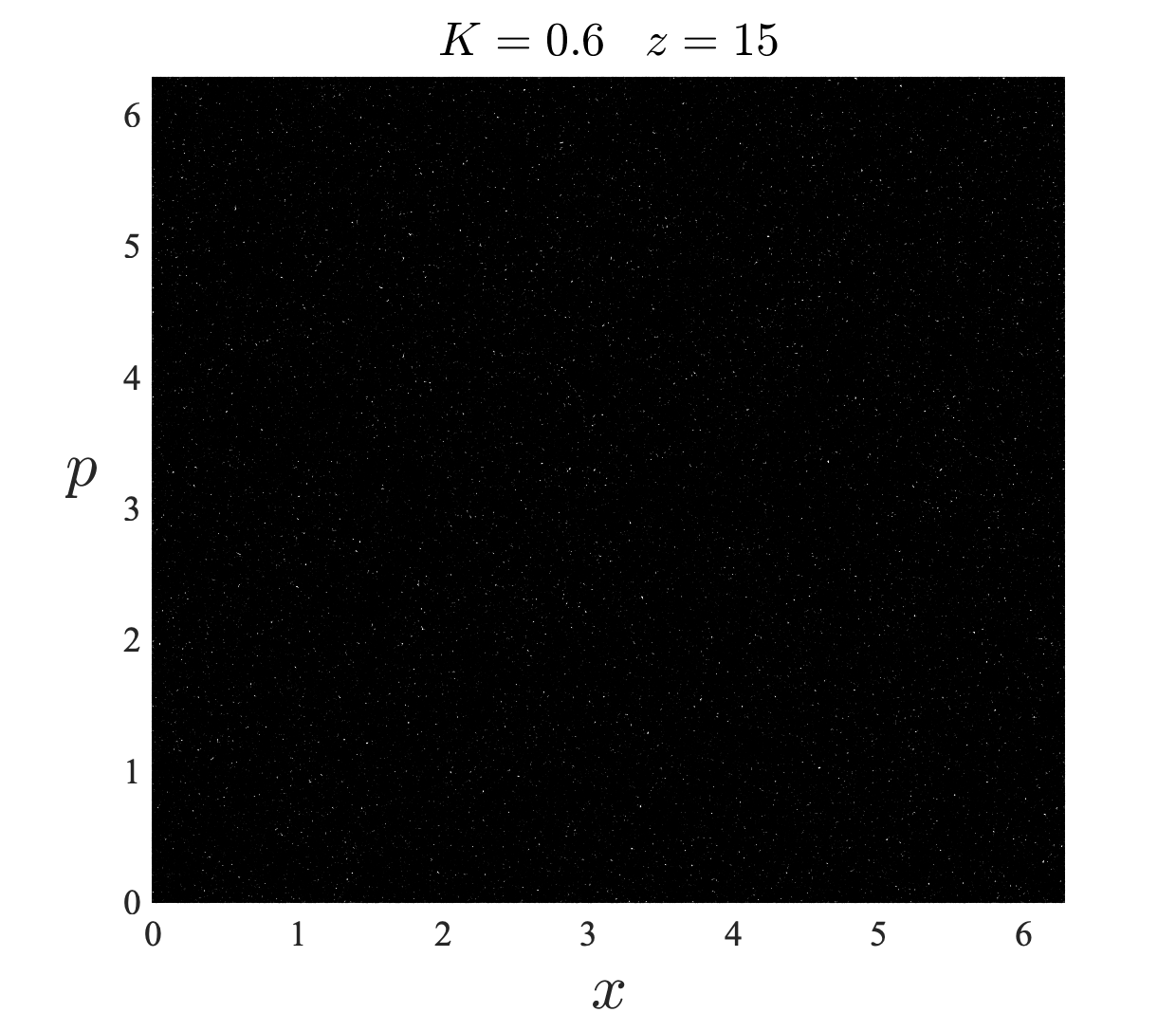}
\includegraphics[height=4.87cm,angle=0]{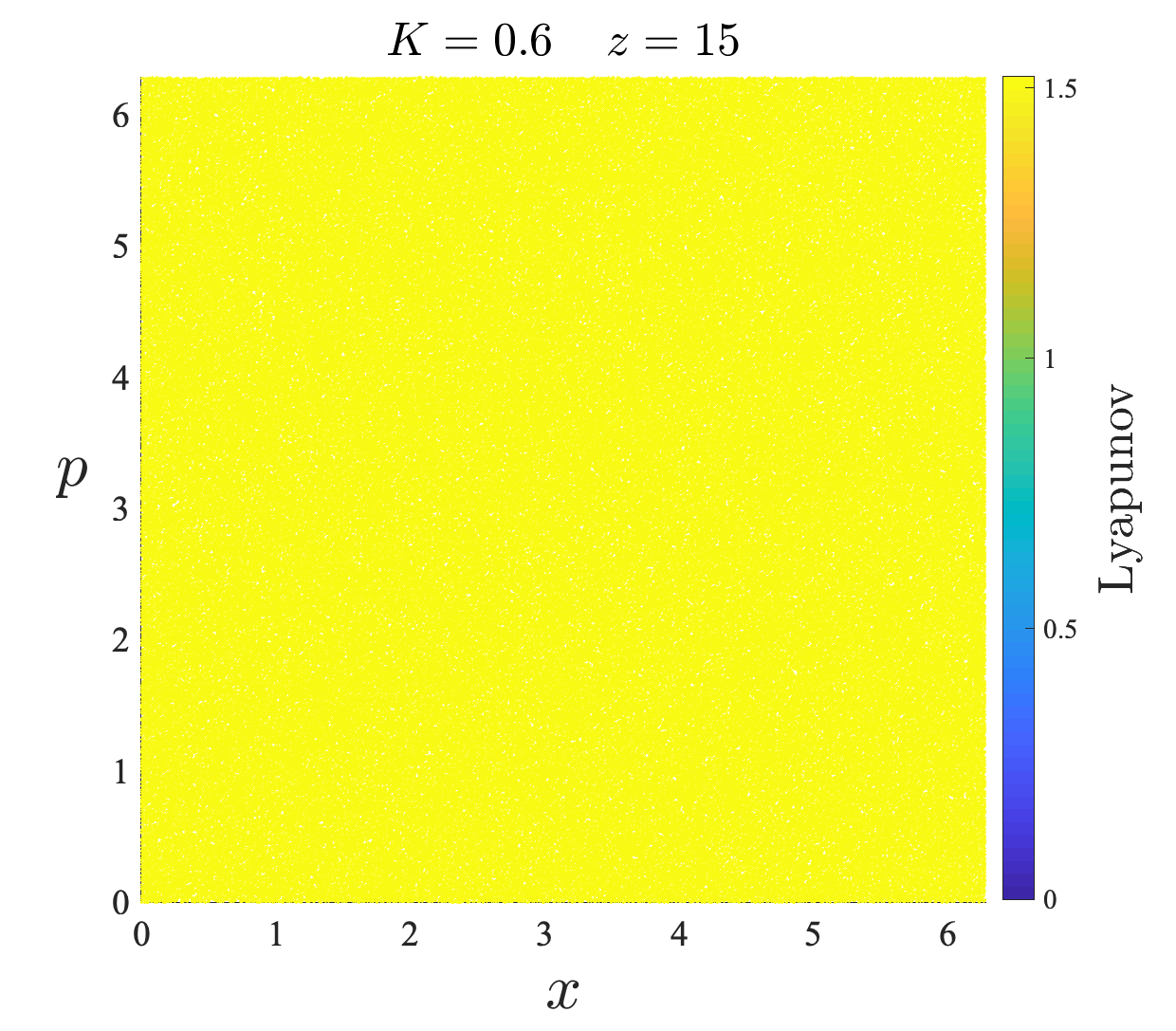}
\includegraphics[height=4.8cm,angle=0]{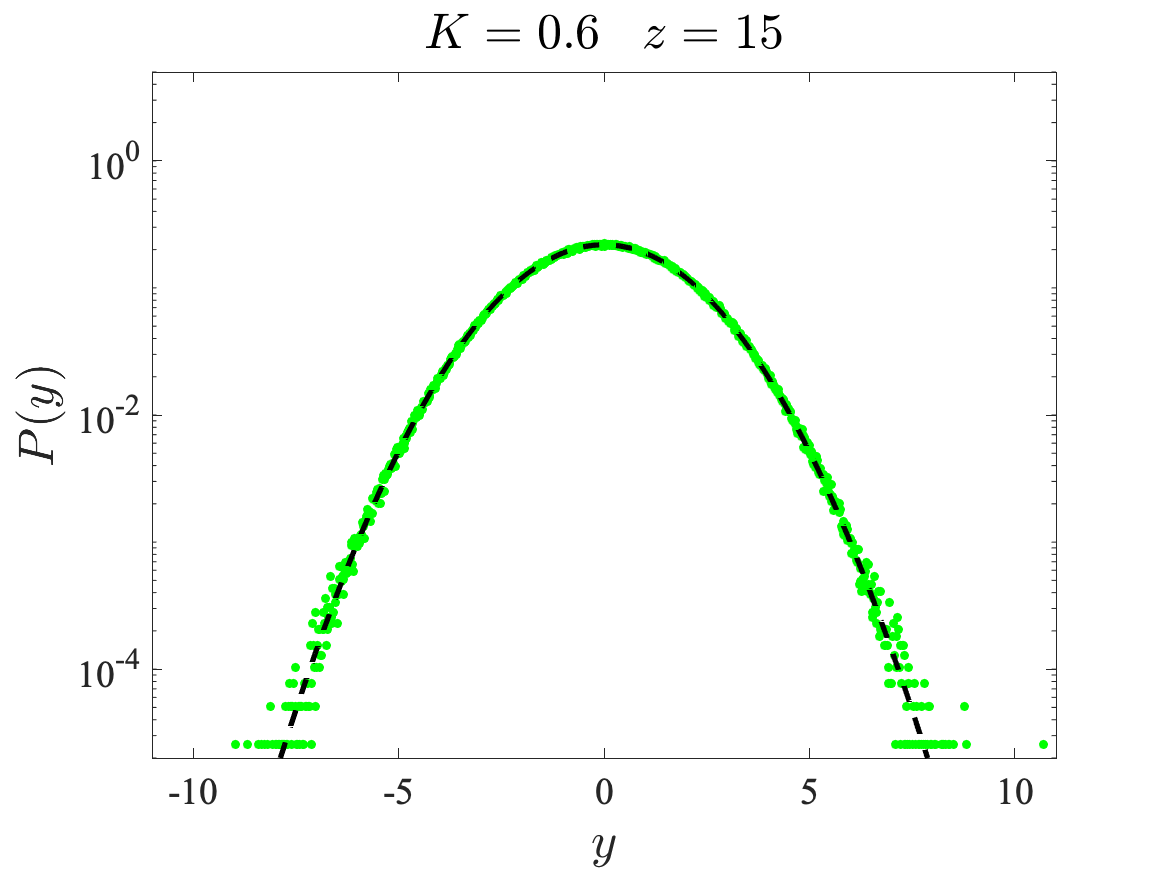}\\
\caption{\label{fig:Fig3} {\small (Color online) 
{\it Left column:}~Phase space portrait of the $z$-standard map for $K=0.6$ with various $z$ values;  
40-50 initial conditions have been used in each case. 
{\it Middle column:}~Lyapunov diagrams of the same cases. 
The Lyapunov exponents have been calculated over 500000 time steps using 500000 initial 
conditions taken randomly from the entire phase space. 
{\it Right column:}~Probability distributions obtained for the same cases. 
In all cases, the number of initial conditions is larger than $3\times 10^{7}$ in order to achieve 
good statistics.}}
\label{Figure3}
\end{figure}

Now, we can concentrate on the $z$-generalized standard map. 
Phase-space portraits of representative cases, their corresponding Lyapunov diagrams, and the  
limit probability distributions are given in Fig.~\ref{Figure2} for $K=0.2$ and in Fig.~\ref{Figure3} 
for $K=0.6$. They enable to visualize how the dynamics of these systems change according to the increment 
of the $z$ term. 
Surprisingly, we noticed that these distributions cannot be modeled using Eq.~\ref{Pq}. 
Instead, we verified that, for this system, the obtained distributions happen to be well approximated 
by  a linear combination of three $q$-Gaussian distributions, namely, 

\begin{equation}
P(y)=\alpha_{q_{1}}P_{q_{1}}(y;\mu_{q_{1}},\sigma_{q_{1}})+
\alpha_{q_{2}}P_{q_{2}}(y;\mu_{q_{2}},\sigma_{q_{2}})+
\alpha_{q_{3}}P_{q_{3}}(y;\mu_{q_{3}},\sigma_{q_{3}}).
\label{Pqq}
\end{equation}
The probability distributions given in Fig.~\ref{Figure2} and Fig.~\ref{Figure3} can be well approximated 
by this linear combination. The relevant parameter values are given in Table.~\ref{Table2}.

\begin{table}[H]
  \caption{Obtained results for the $z$-generalized standard map for representative values of 
  $z$ and $K$.}
\centering
\begin{tabular}{c|c|c||c|c|}
 \cline{2-5}
\multirow{2}{*}{} & 
\multicolumn{2}{|c||}{$K=0.2$} & %
    \multicolumn{2}{|c|}{$K=0.6$} \\
     \cline{2-5}
  & $z=3$ & $z=5$ & $z=3$ & $z=4$  \\
\cline{2-5}
\cline{2-5}
\hline
\hline
\multicolumn{1}{|c||}{$q_1$}& 1.935&  1.935& 1.935& 1.935 \\
\hline
\multicolumn{1}{|c||}{$q_2$}& 1.40&  1.55& 1.40 & 1.55  \\
\hline
\multicolumn{1}{|c||}{$q_3$}& 1&  1.45& 1 &  1 \\
\hline
\multicolumn{1}{|c||}{$\alpha_{q_1}$}& 0.963 & 0.515  & 0.265 & 0.114  \\
\hline
\multicolumn{1}{|c||}{$\alpha_{q_2}$}& 0.025  & 0.340 & 0.565 & 0.300  \\
\hline
\multicolumn{1}{|c||}{$\alpha_{q_3}$}& 0.012 & 0.145  & 0.170 & 0.586 \\
\hline
\multicolumn{1}{|c||}{$B_{q_1}$}& 60000 &  25800 & 25300 & 2000   \\
\hline
\multicolumn{1}{|c||}{$B_{q_2}$}& $0.389$&  $0.072$& $0.0215$ &$0.0039$  \\
\hline
\multicolumn{1}{|c||}{$B_{q_3}$}& $0.00015$&  $0.004$& $0.0455$ &$0.0025$  \\
\hline
\end{tabular}
\label{Table2}
\end{table}

In Fig.~\ref{fig:Fig2}, we see that the phase spaces of $z=1$ and $z=40$ systems are entirely 
occupied by nonergodic stability islands and ergodic chaotic sea, respectively. Conformably with 
these phase space behaviors, the probability distribution of Eq.~\ref{eq:variable} obtained for the 
initial conditions randomly chosen from the entire phase space is well fitted by a Gaussian when 
the system is ergodic-like and by a $q$-Gaussian with $q \simeq 1.935$ when the system is nonergodic-like. 
In Fig.~\ref{fig:Fig3}, for $K=0.6$, the occurrence of a Gaussian as a limit distribution of the initial 
conditions chosen from the ergodic phase space is also verified for $z=15$ system whose phase space is 
occupied by the chaotic sea. 
For $K=0.6$ parameter value of the $z=1$ system which corresponds to the original standard map, 
the phase space consists of both stability islands and the chaotic sea. In accordance with recent 
works \cite{Ugur,Ruiz}, the limit probability distribution of Eq.~\ref{eq:variable}, given in 
Fig.~\ref{fig:Fig3}, is obtained as a linear combination of a Gaussian arises from the initial conditions 
located in the chaotic sea and a $q$-Gaussian with $q= 1.935\pm0.005$ arises from the initial 
conditions located in the stability islands. 
By considering the regions of different behaviors that the 
phase space consists of, These limit distributions are of course expected due to the ergodic/nonergodic 
behavior of the related phase space. 
Here contribution ratios of each term in the linear combination are detected using the Lyapunov spectrum, 
and therefore they are not fitting parameters but determined directly from the dynamics of the system.

We come across with a more complicated limit behavior for the other parameter values of the generalized 
systems with $K=0.2$ and $K=0.6$ cases. Even though the phase spaces of these systems, 
given in Fig.~\ref{fig:Fig2} and Fig.~\ref{fig:Fig3}, consist of both stability islands and the chaotic 
sea similar to the original standard case, the obtained limit distributions exhibit three-component 
behavior which can be modeled by Eq.~\ref{Pqq}. 
For each system mentioned above, relevant parameter values of probability 
distributions, shown in Fig.~\ref{fig:Fig2} and Fig.~\ref{fig:Fig3}, are given in Table.~\ref{Table2}. 
We see that unexpected third component is obtained as a $q$-Gaussian 
with different $q$ values for ($K=0.2$, $z=3$), ($K=0.6$, $z=3$) and ($K=0.6$, $z=4$) cases. 
More surprisingly, obtained probability distribution of ($K=0.2$, $z=5$) case consists of three 
$q$-Gaussians. In order to explain these interesting observations, we have to focus on the 
requirements for the occurrence of the $q$-Gaussians and the phase space behavior of the 
$z$-generalized standard map.

With the present $z$-generalization, we actually create  systems 
with different phase space behavior. As it can be seen from the phase spaces given in 
Fig.~\ref{fig:Fig2} and Fig.~\ref{fig:Fig3}, for typical values of $K$, increment of $z$ term 
increases the amount of nonintegrability of the system and chaotic behavior may occur for 
smaller $K$ values compared to the original standard map case. With the increasing nonintegrability \cite{Hilborn}, 
the stability islands which are actually survived KAM tori dissolve according to the KAM 
theorem \cite{Arnold1978} and the Poincar\'{e}-Birkhoff theorem \cite{Birkhoff1935}. 

According to the Poincar\'{e}-Birkhoff theorem, as a result of the resonances arisen with increased 
nonintegrability, each torus dissolves into alternating sequence of hyperbolic-elliptic points series depending on its winding 
number. Elliptic orbits occur around each elliptic point and in-sets and out-sets of hyperbolic points surround these elliptic orbits. Each elliptic-hyperbolic points series and in-sets and out-sets of hyperbolic points constitute a resonance. As the nonintegrability increases, resonances begin to overlap and destroy surviving tori that were in the region between them \cite{Hilborn}. Chaotic behavior occurs with complex tangle structures created by in-sets and out-sets of the hyperbolic points. Homoclinic and heteroclinic tangles surround the elliptic orbits without intersecting them and a chaotic trajectory spread throughout this tangle structure by exhibiting mixing behavior \cite{Schuster}. Dissolution of tori and enlarging chaotic behavior create archipelagos centered around the elliptic points in the phase space. Outer stability islands of the archipelagos continue to dissolve according to the KAM 
theorem and the Poincar\'{e}-Birkhoff theorem and chaotic behavior occupies larger portion of the phase space. It is important to note that this formation can be used to explain the occurrence of chaotic behavior in Hamiltonian systems and chaotic sea observed in the phase space corresponds to a single constant energy region of the Hamiltonian. By considering the resonance structure discussed above, we can conclude that complex homoclinic and heteroclinic tangles should stick around the elliptic orbits. This sticky behavior exists in the original standard map but chaotic trajectories spend less time in these sticky regions before escaping into the chaotic sea where they can wander throughout apparently randomly. When we modify the sine term in standard map as given in Eq.~\ref{eq:smap}, we radically change the resonance behaviors and alternating sequence of hyperbolic-elliptic points series observed in the original case. As it can be seen from archipelagos in Fig.~\ref{fig:Fig2} and Fig.~\ref{fig:Fig3}, each generalized system has different elliptic-hyperbolic points organization with periodicity related to the $z$ term in sine function. Starting from the integrable case, the case that is common for all $z$ values, tori located in resonance structures dissolve because of the overlaps. Sine term of the generalized standard map affects the number of hyperbolic-elliptic points and their positioning in the phase space and as a result of this effect resonances become stronger with increasing $z$ term for a constant $K$ value. Thus, more in-sets and out-sets of hyperbolic points create more complex tangle structures around stability islands. As these tangles tend to surround and stick around the stability islands, with more complicated structure chaotic trajectories spend more time for covering these tangle structures. Even though these sticky regions are connected to strongly chaotic sea, can be seen clearly from Fig.~\ref{fig:Fig2} and Fig.~\ref{fig:Fig3}, an initial condition 
starting from the sticky region may spend most of its time in that region and may 
escape into the strongly chaotic sea after unpredictable time steps. Statistical effectiveness of sticky regions is presented in the probability distributions obtained for $z>1$ systems, i.e. especially for ($K=0.2$, $z=5$) case which explained in detail below. A chaotic trajectory should visit both chaotic sea and each sticky region, because the chaotic sea 
we see in the phase space is actually an allowed energy surface of the original Hamiltonian system 
which is created by dissolution of the constant energy surface tori. As a chaotic trajectory wanders 
throughout the allowed energy region apparently randomly, its behavior displays a mixing property 
and the system is said to be ergodic in that region. 

When the phase space is fully occupied by the strongly chaotic sea, the whole system is ergodic and 
Gaussian distributions are obtained, which indicates the validity of the BG statistical framework. 
On the contrary, as a consequence of the ergodicity breakdown, $q$-Gaussians are appropriate 
distributions that describe the whole system with phase-space entirely occupied by stability islands. 
As these inferences are verified for the ($K=0.2$, $z=1$), ($K=0.2$, $z=40$) and ($K=0.6$, $z=15$) 
systems, two-component linear combinations of probability distributions of the ($K=0.6$, $z=1$) system 
shows that $q$-Gaussians arise from the initial conditions selected from stability islands whereas the 
Gaussian contribution comes from chaotic trajectories. It is important to note here that ergodicity 
breakdown alone is not sufficient for the occurrence of the $q$-Gaussians; indeed, special type of 
correlations among random variables are also needed. These requirements are fulfilled for stability 
islands of area-preserving maps \cite{Ugur,Ruiz,Ruiz2017b} and, for chaotic bands, of band-splitting 
structure that approaches the chaos threshold of the dissipative logistic map by means of a 
Huberman-Rudnick-like scaling law \cite{ugur2007,ugur2009,Afsar2013}.  

We see from Table~\ref{Table2} that the same $q$-Gaussian contributions with $q= 1.935\pm0.005$
come for each system. By taking into consideration that the same distribution is obtained for initial 
conditions selected from the stability islands of the original standard map ($z=1$) with different 
parameter values \cite{Ugur,Ruiz}, we distinguish stability islands from the phase space by 
using the Lyapunov diagrams given in Fig.~\ref{fig:Fig2} and Fig.~\ref{fig:Fig3}. Moreover, we verify 
that the phase-space occupation ratios of the stability islands of each case are exactly the 
$\alpha_{q_{1}}$ values given in Table~\ref{Table2}. If the initial conditions from the stability 
islands are discarded, we are left with the chaotic sea in the phase space which should be related to 
the remaining components of the probability distribution given in Eq.~\ref{Pqq}. 
Let us start by analyzing the ($K=0.2$, $z=3$), ($K=0.6$, $z=3$) 
and ($K=0.6$, $z=4$) cases whose chaotic seas give rise to both a Gaussian and a $q$-Gaussian. 
When we look at the chaotic seas in the phase spaces of these systems in more detail, we see that 
strongly sticky regions which cannot occur in the original standard map arise for all these systems. 
Due to the resonance behaviors mentioned above, these sticky regions surround archipelagos and create 
complicated structures which are also connected with the chaotic sea. An initial condition located 
inside one of these sticky regions may give rise to a chaotic trajectory that stays inside this region 
for many iteration steps. 
An initial condition inside the sticky region evolves to cover this region, after unpredictable iteration 
steps, this trajectory may escape to the chaotic sea and may wander throughout the sea apparently 
randomly. In the probability distribution analyses, we use large but finite iteration steps and it seems 
that during this iteration steps chaotic trajectories located inside the sticky regions cannot display 
strong mixing behavior by confining inside these regions and not visiting large portion of the allowed 
energy region. Chaotic trajectories that do not 
enter sticky regions for large iteration steps can freely wander in the large portion of the allowed 
energy region and give rise to a Gaussian distribution. 
In principle, if it was possible to leave the system to evolve infinitely, the entire equal energy 
region would be covered by a single chaotic trajectory. 
In our observation interval, as some of the chaotic trajectories wander freely, some of them instead spend 
most of their times in the sticky region. Most probably, this observation may be a plausible explanation 
of the obtained limit probability distributions of systems that we investigate here. 
Observations made for the ($K=0.2$, $z=5$) case present more complicated scenario. 
When we look at the Lyapunov spectrum of this case in Fig.~\ref{fig:Fig2}, we see the horizontal 
band-like structures. Based on our observations, the chaotic sea of each band acts as a sticky region 
in the phase space and all these regions are connected by not having any stability island as a barrier between 
them. When we analyze the phase space 
behavior of the chaotic trajectories by selecting initial conditions inside the chaotic seas and 
letting them to evolve, we observe that chaotic trajectories do not spread into the allowed energy 
region randomly as expected from the regular chaotic trajectory like we see in ($K=0.2$, $z=40$) and 
($K=0.6$, $z=15$) systems. 
Instead of spreading into the allowed energy region randomly, chaotic trajectories first cover their 
bands' chaotic sea and then move into another band. In our observations we see that even for large 
number of iteration steps, i.e. $T=10^{10}$, entire chaotic sea cannot be covered by a trajectory 
starting from an initial condition. As chaotic trajectories move in the phase space covering firstly 
one band and then the others respectively, mixing property of this system is completely different from 
that of the original system. This difference seems to be related to the occurrence of two $q$-Gaussian 
contributions detected for the limit probability distribution from the chaotic sea.

\begin{figure}[H]
	\centering
	\includegraphics[scale=0.5,angle=0]{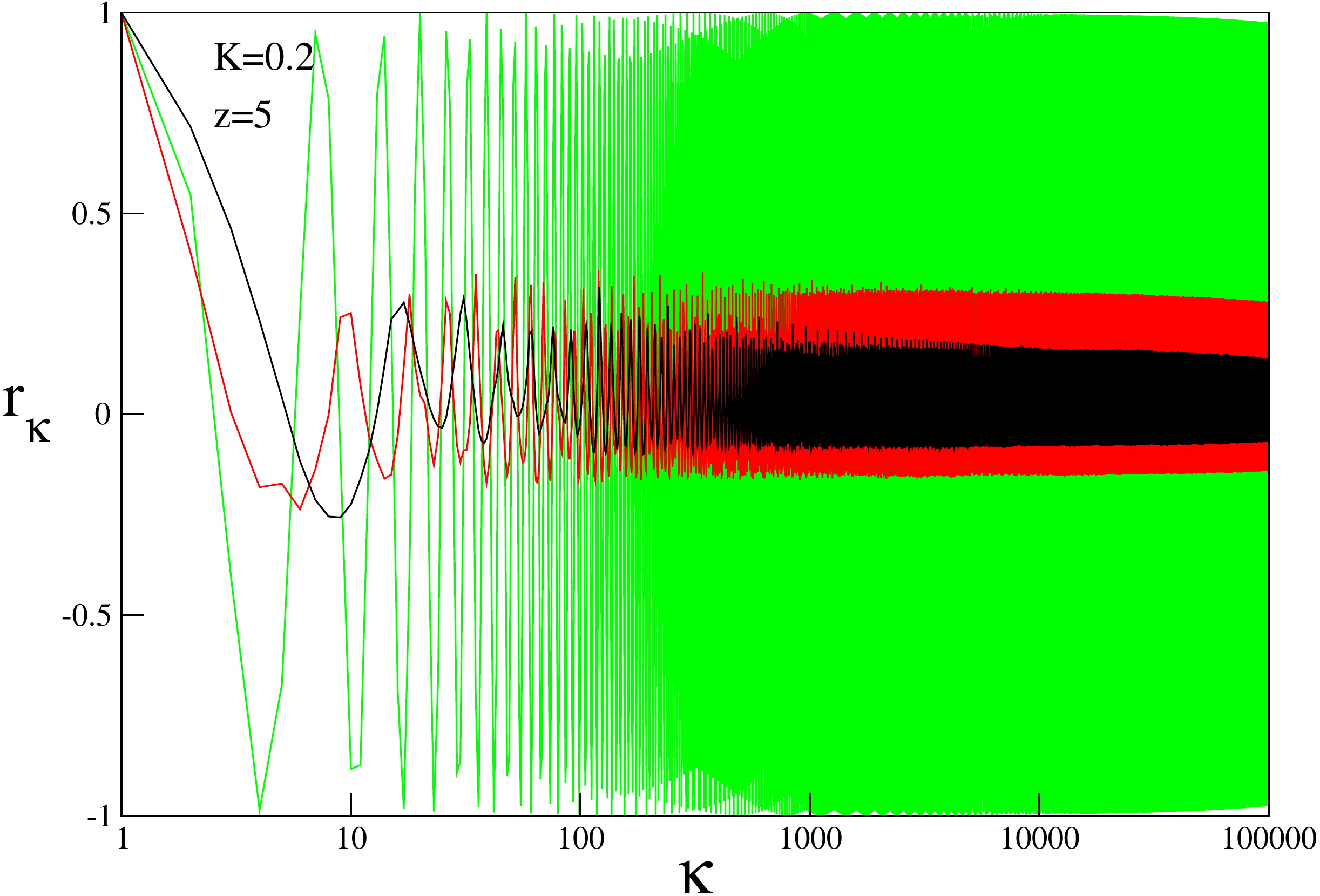}
	\caption{\label{fig:Fig4} {\small (Color online) 
	Three types of auto-correlation functions obtained for the $K=0.2$, $z=5$ case. 
	The green curve is a representative result for a trajectory produced from an 
	initial condition taken from the stability island (corresponding to a $q$-Gaussian with $q= 1.935\pm0.005$). 
	The red and black curves are two representative results from trajectories started from the chaotic sea 
	(corresponding to Gaussians with different $B$ values). 
	Nonvanishing character of all auto-correlation functions is evident here. 
	}}
\end{figure}

\begin{figure}[H]
\centering
\includegraphics[scale=0.5,angle=0]{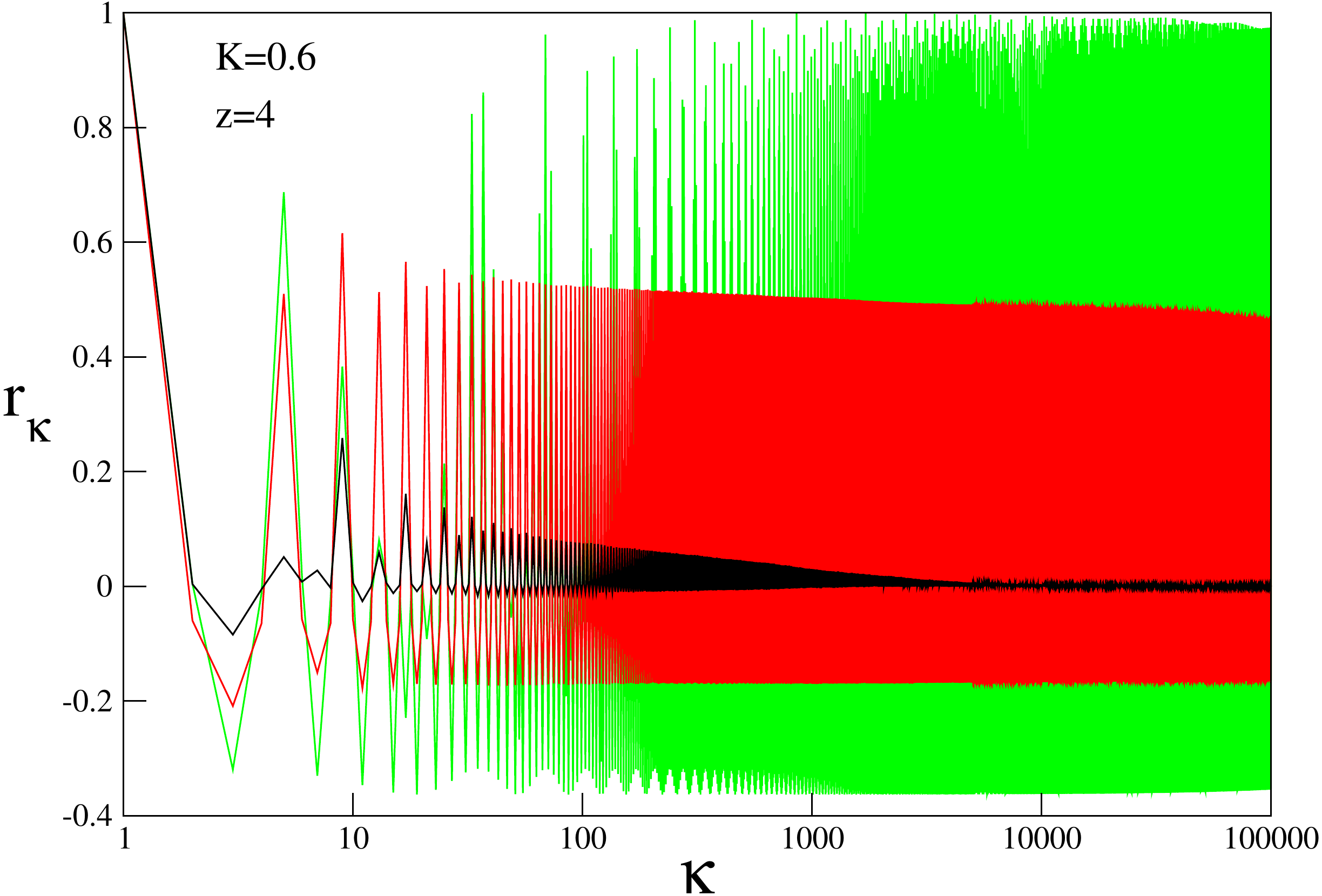}
\caption{\label{fig:Fig5} {\small (Color online) 
Three types of auto-correlation functions obtained for the $K=0.6$, $z=4$ case. 
The green curve is a representative result for a trajectory produced from an 
initial condition taken from the stability island (corresponding to a $q$-Gaussian with $q= 1.935\pm0.005$). 
The red and black curves are two representative results from trajectories started from the chaotic sea 
(corresponding to a $q$-Gaussian with $q\simeq1.61$ and a Gaussian, respectively). 
Nonvanishing character of auto-correlation function is seen for one of 
them (red curve) which belongs to a trajectory of an initial condition that spends 
considerable time in the sticky region. Vanishing of the auto-correlation function 
for an initial condition (black curve) that does not enter the sticky regions is also 
evident. 
   }}
\end{figure}

In order to improve our understanding of the emergence of the $q$-Gaussians, we have also analyzed
the auto-correlation function $r_{\kappa}$ 
defined as follows
\begin{equation}
r_{\kappa}=\sum_{i=1}^{T-\kappa}\frac{(y_{i}-\langle y\rangle)(y_{i+\kappa}-\langle y\rangle)}
{\sum_{i=1}^{T}(y_{i}-\langle y\rangle)^{2}}
\label{autocorrelation}
\end{equation}
where $\kappa$ is the time lag, $T$ is the number of iteration steps which constitute the trajectory,
and $\langle y\rangle=T^{-1}\sum_{i=1}^{T}y_{i}$ \cite{MoralesNuevo1993}. Iterates of a trajectory 
are correlated for $r_{\kappa}\neq 0$ and not correlated for $r_{\kappa}=0$. For each system 
given in Table~\ref{Table2}, we randomly choose a large number of initial conditions from the entire 
phase space and let the system to evolve along $T=2^{22}$ iteration steps, which coincides with the 
iteration number used in the probability distribution computations, starting from these initial 
conditions. In the computations of $r_{\kappa}$, for each trajectory, we use $\kappa=10^{5}$ as a 
maximum time lag by considering large computation times required for larger time lag values. 
Obtained results show that all systems exhibit three different tendencies for the auto-correlation 
function compatible with the probability distributions in the form of Eq.~\ref{Pqq}. 
In Fig.~\ref{fig:Fig4} and Fig.~\ref{fig:Fig5} we demonstrate the auto-correlation functions of the
($K=0.2$, $z=5$) and ($K=0.6$, $z=4$) systems as a function of the logarithm of the time lag 
respectively to corroborate the explanations given for nonergodic and nonmixing behavior of 
the chaotic trajectories. The logarithm of the time lag is used and $\kappa$ is cut at $10^{5}$ 
in order to obtain a better visualization of the oscillatory behavior of the auto-correlation functions. 
In both figures common green color is used to indicate the auto-correlation functions of iterates 
starting from initial conditions located inside one of the stability islands, 
e.g. ($x=5.042890762...$, $p=0.154394798...$­) for ($K=0.2$, $z=5$) system and 
($x=1.617417044...$, $p=4.837677782...$­) for($K=0.6$, $z=4$) system. 
As seen from figures, green curve oscillates around zero with very large amplitudes 
and this behavior indicates that the iterates in the stability islands are strongly correlated. 
In Fig.~\ref{fig:Fig4} we see that red and black curves also oscillate around zero with 
different amplitudes that are smaller than the amplitude of the green one. 
Both of these functions are obtained for trajectories whose initial conditions are selected from 
the chaotic sea of the phase space, 
i.e. ($x=3.190574110......$, $p=7.259267286...\times 10^{-2}...$) for red and 
($x=3.096157726...$, $p=9.095801865...\times 10^{-2}...$) for black. 
These auto-correlation function behaviors are three main types of correlations that are observed 
in the analyses of the ($K=0.2$, $z=5$) system and through this observations we can say that the 
chaotic trajectories in the phase space exhibit correlated behavior which is weaker than the 
correlation among the iterates of a trajectory in the stability islands. These three type of 
correlations observed in the phase space together with the nonergodicity of the present system 
mentioned above fulfill the requirements of the occurrence of the $q$-Gaussians and are thought to 
explain the obtained limit probability distribution which is a linear combination of three $q$-Gaussians. 
In Fig.~\ref{fig:Fig5} red and black curves are obtained for the initial conditions chosen 
from the chaotic region in the phase space. 
As initial condition ($x=5.654566893...$, $p=1.627640289...$) of red curve is 
located in the sticky region, ($x=5.668539896...$, $p=4.509105458...$) initial condition 
of black curve is located in the strongly chaotic sea. When we look at the figure, 
we see that black curve decreases to zero after a short time lag and oscillates around 
zero by indicating uncorrelated nature of the iterates of the trajectory. This auto-correlation function 
behavior is similar to the auto-correlation function obtained for the white noise which recently 
shown in Ref.\cite{Cetin2015}. On the contrary, red curve oscillates around zero with 
large amplitudes like the previous scenario. From these observations we can deduce that chaotic 
trajectories located inside the sticky regions may show correlations and chaotic trajectories that 
do not enter into the sticky regions for a long period of iteration steps display expected uncorrelated 
behavior of the chaotic trajectories. Also, we obtain the same auto-correlation function behaviors for 
($K=0.2$, $z=3$) and ($K=0.6$, $z=3$) systems that exhibits similar limit probability distribution 
like ($K=0.6$, $z=4$) system. These observations seem to provide an adequate explanation for the obtained 
limit probability distributions.

\section{Conclusions}

In this work, our results on the area-preserving maps can be summarized by classifying them into two 
groups: observations for the stability islands and for the chaotic trajectories. 
For the Harper map and several $z$-generalized standard map systems, for large number 
of iteration steps, the limit probability distributions coming from the sum of the iterates when the 
system is started from initial conditions located inside the stability islands seem to 
converge to a $q$-Gaussian with $q= 1.935\pm0.005$ value. 
Each different $z$ value corresponds to a new area-preserving system. 
Taking into account all of the maps analyzed in this paper and also results of the recent paper on the 
statistical characterization of the area-preserving web map \cite{Ruiz2017b}, the main goal of this 
manuscript is to verify numerically that the limit probability distribution obtained when the system is 
initiated from initial conditions located in the stability islands is always well approximated by a 
$q$-Gaussian with $q= 1.935\pm0.005$ value. 
Regardless of the magnitude of the phase space occupation ratios of the stability islands 
(these ratios are not fitting parameters since they come directly from the Lyapunov spectrum of the 
system) for various 
map parameter values, $q$-Gaussian with $q= 1.935\pm0.005$ maintains its presence together with other 
distributions and this fact indicates that $q$-Gaussian with $q= 1.935\pm0.005$ value is a robust limit 
behavior for the stability islands of the area-preserving maps. 

Although the stability islands of the area-preserving maps exhibit the same limit behavior, unexpected 
observations are made for chaotic trajectories of the different maps. Considering the definitional 
properties of the chaotic trajectories, e.g. the apparently random behavior and the exponential 
divergence of initially nearby trajectories, one can suggest that the chaotic trajectories 
wander freely through the allowed energy region and they spread into this region by displaying the 
mixing property. Under normal circumstances, a single chaotic trajectory rapidly spreads into 
allowed region and outlines this region after a few iteration steps. This common behavior of the 
chaotic trajectories is observed for all control parameter values of the Harper map. 
For all cases of the Harper map, when the chaotic region develops, we obtain Gaussian distributions 
for the limit behavior of the sums of the iterates for the initial conditions started from the chaotic 
sea, as expected. 
However, as shown here, some chaotic trajectories cannot exhibit regular behavior of the 
chaotic trajectories due to the sticky regions occur around stability islands for some 
area-preserving maps.  
Such a chaotic trajectory may not visit most of the allowed energy region during the 
observation time and therefore it cannot behave similarly as the chaotic trajectories which do not 
visit sticky regions during the same period of time. 
Even though chaotic trajectories exponentially diverge while covering the sticky region, magnitude 
of their divergence is much smaller compared to the divergence in the chaotic sea as seen from the 
Lyapunov spectra.  
A second $q$-Gaussian is obtained in limit 
probability distributions of systems that exhibit sticky behavior in their phase spaces and this might be 
explained due to different mixing property compared to the standard case and correlated nature of 
chaotic trajectories of sticky region. 
Even though contribution ratios of these second $q$-Gaussians cannot be determined directly from the 
Lyapunov spectra as we did before, they are thought to be as robust as other distributions that make 
contributions to the limit behavior for long but finite time intervals.

\section*{Acknowledgments}
We acknowledge fruitful remarks from two anonymous Referees. 
This work has been partially supported by TUBITAK (Turkish Agency) under the 
Research Project number 115F492, and by CNPq and Faperj (Brazilian Agencies).


\end{document}